%% file: lrreadout.tex
\documentclass[groupedaddress,twocolumn,10pt,prl,aps,showpacs,longbibliography,floatfix]{revtex4-2}

\usepackage[english]{babel}

\usepackage{graphicx}
\usepackage{dcolumn}
\usepackage{bm}
\usepackage{physics}
\usepackage{xcolor}
\usepackage{hyperref}
\usepackage{minitoc}
\newcommand\LL{\mathcal{L}}

\begin{document}

\preprint{APS/123-QED}

\title{Low-rank optimal control of quantum devices}

\author{Leo Goutte}
\email{leo.goutte@epfl.ch}
\author{Vincenzo Savona}%
\email{vincenzo.savona@epfl.ch}
\affiliation{Institute of Physics, Ecole Polytechnique Fédérale de Lausanne (EPFL), CH-1015 Lausanne, Switzerland}
\affiliation{Center for Quantum Science and Engineering, Ecole Polytechnique Fédérale de Lausanne (EPFL), CH-1015 Lausanne, Switzerland}
\date{\today}

\begin{abstract}

We demonstrate that the control protocols of quantum information devices can be simulated by assuming a low-rank ansatz for the density matrix. The rationale underlying this assumption is that quantum information protocols, by design, operate in a regime of nearly pure quantum states. Within the low-rank assumption, the simulation of these protocols is considerably faster than solving the full Lindblad master equation. This advantage can be used to increase the accuracy of the simulation by avoiding uncontrolled approximations, and to streamline protocol optimization. We benchmark our approach on the optimization of the transmon qubit dispersive readout in a realistic transmon-resonator-filter model. With Hilbert space dimension $N = 2000$, assuming a rank as low as $M = 20$ we achieve a nearly 100-fold speedup compared to full master equation integration while accurately reproducing all relevant observables. By combining the low-rank approximation with a compact pulse parametrization and gradient-free optimization, we obtain state-of-the-art readout assignment errors $\varepsilon_a \approx 1.2 \times 10^{-3}$ for a 40 ns readout pulse schedule, while comfortably running on a laptop and not relying on the rotating-wave approximation. Our approach is broadly applicable to most quantum control protocols, including quantum gates, state preparation, and fast reset operations. This establishes low-rank methods as a general tool for optimal control across diverse quantum platforms.


\end{abstract}

\maketitle


\emph{Introduction.---} Quantum computing promises transformative advances in quantum chemistry, materials science, and optimization, offering dramatic speed-ups for specific problem classes. Realizing this potential demands solving major hardware and software challenges, particularly in the precise control, design, and optimization of quantum devices.

Quantum optimal control \cite{peirce_optimal_1988,werschnik_quantum_2007,Machnes_comparing_2011,koch_quantum_2022,koch_controlling_2016,schmidt_optimal_2011} tackles these challenges by designing control schedules for executing elementary operations such as quantum gates \cite{sauvage_optimal_2022,sarma_designing_2025,jandura_time-optimal_2022,schulte-herbrueggen_optimal_2011,werninghaus_leakage_2021,gautier_designing_2023,propson_robust_2022,blumel_efficient_2021,rahman_learning_2024,ding_high-fidelity_2023,hyyppa_reducing_2024,sung_realization_2021,tripathi_suppression_2022,egger_optimized_2013,blumel_power-optimal_2021,mohan_robust_2023,kuzmanovic_neural-network-based_2025,genois_quantum_2024, nam_nguyen_reinforcement_2024}, quantum bit readout \cite{heinsoo_rapid_2018, gautier_optimal_2025, chen_fast_2024, abdelhafez_gradient-based_2019, gambetta_protocols_2007, kurilovich_high-frequency_2025, chapple_balanced_2025, swiadek_enhancing_2024, chen_transmon_2023, bengtsson_model-based_2024, chatterjee_enhanced_2025}, and state reset \cite{chen_fast_2024,abdelhafez_gradient-based_2019,gautier_optimal_2025,egger_pulsed_2018,mcewen_removing_2021}. At larger scales, the simultaneous control of many interacting quantum bits requires the orchestration of numerous control pulses and frequencies \cite{bengtsson_model-based_2024, klimov_optimizing_2024, klimov_snake_2020, sivak_real-time_2023, acharya_quantum_2025, sivak_optimization_2024, acharya_suppressing_2023}. These optimization tasks typically involve numerous evaluations of a loss function that quantifies the performance of the control protocol and depends on a large number of parameters, including pulse shapes, durations, and frequencies. The loss function is then typically evaluated through the accurate simulation of the quantum device dynamics. 
State-of-the-art hardware quantum platforms based on superconducting circuits, trapped ions, Rydberg atoms, are all \emph{open quantum systems}, where the dissipation and decoherence significantly affect the system dynamics. Within the assumption of a Markovian environment, these dynamics are modeled by the Lindblad master equation~\cite{breuer_theory_2007}, whose computational complexity scales as $\mathcal{O}(N^2)$. Accurate and predictive simulations often require large values for the Hilbert space dimension $N$, making the optimization process prohibitively expensive.

To address these challenges, significant effort has been devoted to developing efficient optimization methods~\cite{Khaneja_optimal_2005,Krotov1993,Goerz_Krotov_2019,Caneva_chopped_2011,muller_information_2022,gautier_optimal_2025,leung_speedup_2017,abdelhafez_gradient-based_2019,schulte-herbrueggen_optimal_2011,Boutin_resonator_2017,Egger_optimal_2014,Goerz_Krotov_2019}. However, less effort has been directed toward reducing the computational time required to solve the Lindblad master equation itself through controlled approximations that maintain predictive accuracy of the optimization metrics.

In this letter, we present a computational framework for quantum optimal control that enables efficient, accurate, and scalable optimization for quantum devices. 
The central feature is a dramatic speedup in the numerical calculation of the loss function, which we achieve by adopting a Low-Rank Approximation (LRA)~\cite{lebris_lowrank_2013,mccaul_fast_2021,chen_low-rank_2021,donatella_continuous-time_2021,santos_low-rank_2025,joubert-doriol_non-stochastic_2014,joubert-doriol_problem-free_2015,gravina_adaptive_2024}. The LRA is successful because, in most quantum operation protocols, the system remains in a nearly pure quantum state throughout the operation schedule. In such regimes, the density matrix of the system contains much less information than implied by its full $N \times N$ dimensionality, and can be represented with rank-$M$ matrices (where $M \ll N$ in high-purity regimes). We benchmark our approach by \emph{deliberately revisiting} the optimization of the dispersive readout of a transmon qubit coupled to a readout resonator and a Purcell filter~\cite{blais_cavity_2004, chen_transmon_2023, swiadek_enhancing_2024, touzard_gated_2019, walter_rapid_2017, sunada_fast_2022, reed_fast_2010, blais_circuit_2021, kurilovich_high-frequency_2025} that was very recently studied by Gautier et al.~\cite{gautier_optimal_2025}. By further combining the LRA with a gradient-free heuristic for the optimization, our approach results in an optimal readout assignment error of $1.2 \times 10^{-3}$, on par with that obtained in Ref.~\cite{gautier_optimal_2025}, while leading to a nearly $100$-times reduction in the computational time required for realistic device simulations.

\begin{figure}
    \centering
    \includegraphics[width=0.9\linewidth]{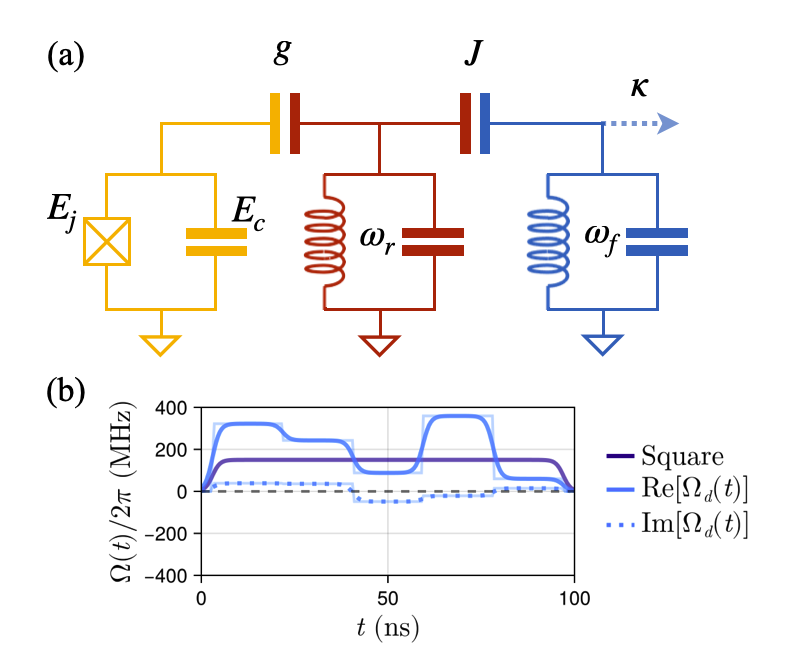}
    \caption{(a) Schematic of the transmon readout model. The transmon (yellow) is capacitively coupled to a resonator (red), itself coupled to a Purcell filter (blue). The filter is coupled to the environment through a readout line that enables driving and introduces dissipation at a rate $\kappa$. (b) Illustrative example of driving pulses $\Omega_d(t)$. Both the square (dark blue) and stepwise (light blue) pulses are smoothed out using a logistic function, as detailed in Ref.~\cite{SM}. The thin lines show the stepwise pulse before smoothing.}
    \label{fig:schematic}
\end{figure}

\emph{Transmon readout model.---}
We model the transmon dispersive readout protocol for the setup considered in Ref.~\cite{gautier_optimal_2025} and illustrated in Fig.~\ref{fig:schematic}(a). The system is described by the Hamiltonian
\begin{align}
\label{eq:hamiltonianpurcell}
\begin{split}
   \hat H =& 4E_c \hat n_t^2 - E_j \cos\hat\varphi_t + \omega_r \hat a^{\dagger}\hat a + \omega_f \hat f^{\dagger}\hat f 
    \\ &-J\left(\hat a^{\dagger} - \hat a\right)\left(\hat f^{\dagger} - \hat f\right) + ig\hat n_t\left(\hat a^\dagger - \hat a\right) \\
    &+ i\Omega_d(t) \sin(\omega_d t)\left(\hat f^\dagger - \hat f\right),
\end{split}
\end{align}
and is governed by a Lindblad master equation
\begin{equation}
\label{eq:hamiltonian}
    \dot{\hat{\rho}} = -i \left[\hat H,\hat\rho\right] + {\kappa} \mathcal{D}\left[\hat f\right]\hat\rho + {\gamma} \mathcal{D}\left[\hat b\right]\hat\rho\,,
\end{equation}
where $\mathcal{D}[\hat L_k]\hat\rho = \hat L_k \hat\rho \hat L_k^\dagger - \{\hat L_k^\dagger \hat L_k, \hat\rho\} / 2$. Here, $\hat n_t$ and $\hat \varphi_t$ are the transmon conjugate charge and phase operators, and $E_c$ and $E_j$ the corresponding charging and Josephson energies. The operators $\hat b$, $\hat a$, and $\hat f$ are annihilation operator of the transmon, resonator, and filter modes respectively, while $\omega_r$ and $\omega_f$ are the bare resonator and filter resonant frequencies. The capacitive coupling strengths are $g$ and $J$, and the dissipation rates are $\kappa$ for the filter and $\gamma$ for the transmon. The system is driven through the filter mode by a microwave pulse at frequency $\omega_d$ with envelope $\Omega_d(t)$. We initially model the latter as a square pulse of maximum amplitude $|\Omega_d|$ with a smooth ramp up and ramp down modulated by a logistic function, as plotted in Fig.~\ref{fig:schematic}b. Importantly, the chosen ramp time of $2.5$ ns is slow compared to $\omega_d^{-1} \approx 0.02$ ns so as to avoid spurious photon populations in the driven mode. 

\begin{figure}
    \centering
    \includegraphics[width=0.9\linewidth]{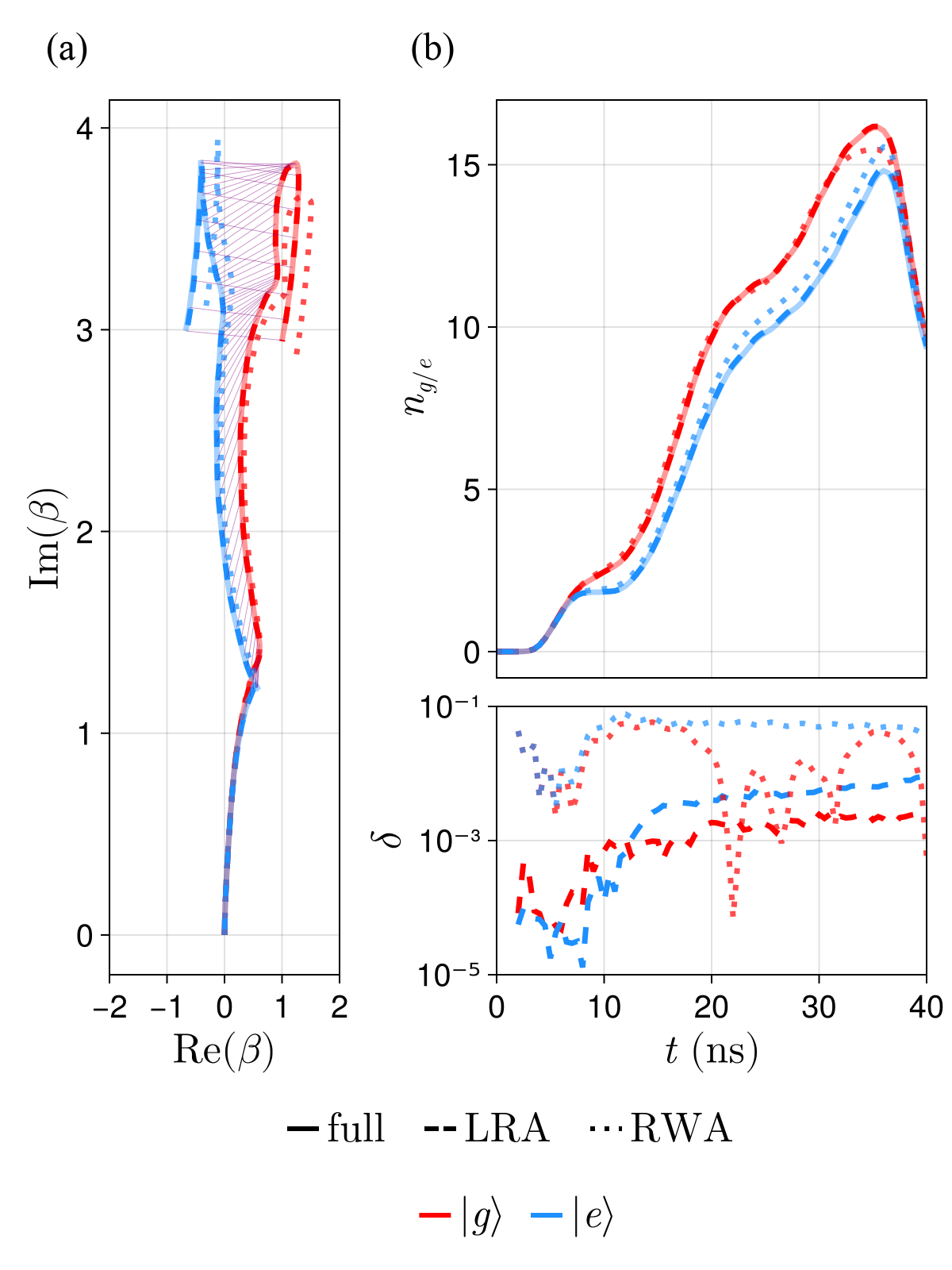}
    \caption{(a) The phase-space trajectories $\beta_{g/e} = \mathrm{tr}(\hat\rho_{g/e} \hat f)$ for a readout time $\tau = 40$ ns and a square pulse profile with amplitude $|\Omega_d| / 2\pi = 150$ MHz. The full, low-rank with $M=20$, and RWA evolutions correspond to the full, dashed, and dotted lines, respectively. The red (blue) lines correspond to an initial ground (first excited) transmon state. The thin lines connecting red and blue trajectories highlight equal times on the two curves. (b) The photon occupation $n_{g/e} = \mathrm{tr}(\rho_{g/e} \hat f^\dagger f) = |\beta_{g/e}|^2$ in the filter mode and the relative errors between the full and low-rank (dashed), and full and RWA (dotted) results. The relative error between the full and approximate values, defined for a quantity $y$ as $\delta=|(y_{\mathrm{full}} - y_{\mathrm{LRA}})/y_{\mathrm{full}}|$ is much smaller for the LRA than the RWA. The Hilbert space truncations are $N_l = 100$, $N_u = 4$, and $N_t = 5$, resulting in $N = 2000$ states. }
    \label{fig:phase_space_trajectories}
\end{figure}

Due to their coupling, the resonator's frequency undergoes a small dispersive shift which depends on the state of the transmon. This in turn affects the resonator dynamics, allowing one to infer the initial state of the transmon from a heterodyne measurement of the resonator field. The addition of a Purcell filter coupled to the resonator does not fundamentally change this picture, but simply suppresses the spontaneous transmon decay during the readout schedule~\cite{walter_rapid_2017,sunada_fast_2022, reed_fast_2010}. The coupled resonator-filter subsystem admits two normal modes and either one of these modes can be used for the readout by resonantly driving the filter at the corresponding hybridized normal-mode frequency~\cite{swiadek_enhancing_2024}. A detailed description of the readout protocol and its numerical implementation is presented in Ref.~\cite{SM}.

\emph{Computational performance.---}
In Fig.~\ref{fig:phase_space_trajectories}, we compare the phase-space trajectories of the field in the filter mode and the corresponding photon occupation as obtained with the full master equation, with the LRA, and with the RWA, for a readout time $\tau = 40$ ns and a square drive. This reproduces the square-pulse setting studied in Ref.~\cite{gautier_optimal_2025}. 

The LRA is applied by truncating the density matrix to a rank $M \ll N$ such that $\hat\rho \simeq \mathbf{m}\mathbf{m}^\dagger$~\cite{joubert-doriol_non-stochastic_2014, joubert-doriol_problem-free_2015, donatella_continuous-time_2021, mccaul_fast_2021, chen_low-rank_2021, santos_low-rank_2025, gravina_adaptive_2024}. Here, $\mathbf m$ is a $N \times M$ matrix containing the low-rank states on each of its columns. Its time evolution is governed by a non-stochastic Schrödinger equation \cite{joubert-doriol_non-stochastic_2014} which is numerically integrated with an adaptive-timestep ODE solver. By representing the density matrix with rank-$M$ matrices, the number of coupled differential equations to solve is reduced from $N^2$ to $MN$, resulting in dramatic savings in memory and computational time. A more detailed derivation and discussion of the LRA's validity can be found in Ref.~\cite{SM}.

Already with a modest rank of $M=20$, the LRA result accurately describes the phase space trajectories for both choices of initial transmon state. The RWA on the other hand results in sizable quantitative differences. For this particular choice of pulse shape and amplitude, this discrepancy still leads to a fairly accurate prediction of the readout assignment error. A full optimization schedule, however, may explore regions in parameter space where the discrepancy brought by the RWA may be larger and possibly lead to sub-optimal results. If this issue can be safely excluded by restricting the parameter space in the optimization, then it is possible to simulate the system dynamics by applying both the LRA and RWA, leading to even faster simulation times. Tab.~\ref{tab:runtimes} shows the wall-clock compute times for the full simulation, the LRA, the RWA, and the two combined. The LRA and combined LRA + RWA results were obtained using a fixed-rank non-stochastic matrix Schrödinger equation solver publicly available through Ref.~\cite{goutte_github_2025}. The LRA with $M = 20$ achieves a $81$-times speedup with respect to the full simulation. This should be compared to the $4$-times speedup brought by the RWA. Combining the LRA with $M=20$ and the RWA, a speedup exceeding $100$-times is obtained. 

\begin{table}
\begin{ruledtabular}
    \begin{tabular}{c|cccccc}
         Model & Full & RWA & LRA & LRA + RWA\\
         \hline
        $\ket{g}$ & 2441 & 623 & 30 & 21 \\
        $\ket{e}$ & 2837 & 741 & 33 & 20
        
    \end{tabular}
\end{ruledtabular}
\caption{Runtime in seconds for the full master equation, RWA, and low-rank solvers with $M=20$ for the transmon readout with a square pulse of $\tau=40~\mathrm{ns}$ duration. The rows correspond to initializing the transmon in the $\ket{g}$ or $\ket{e}$ state, respectively. For the full master equation, we use the \texttt{mesolve} solver from QuantumToolbox.jl~\cite{mercurio_quantumtoolboxjl_2025}. The total dimension of the Hilbert space is $N = 2000$ ($N_l = 100$, $N_u = 4$, $N_t = 5$). The times were obtained from CPU calculations on a laptop with an Apple M3 Pro CPU and 36 GB of RAM. We encourage the interested reader to reproduce these and all other results herein with the help of the QuantumToolbox.jl package~\cite{mercurio_quantumtoolboxjl_2025, goutte_github_2025}.\label{tab:runtimes}}
\end{table}

\begin{figure*}
    \centering
    \includegraphics[width=0.8\linewidth]{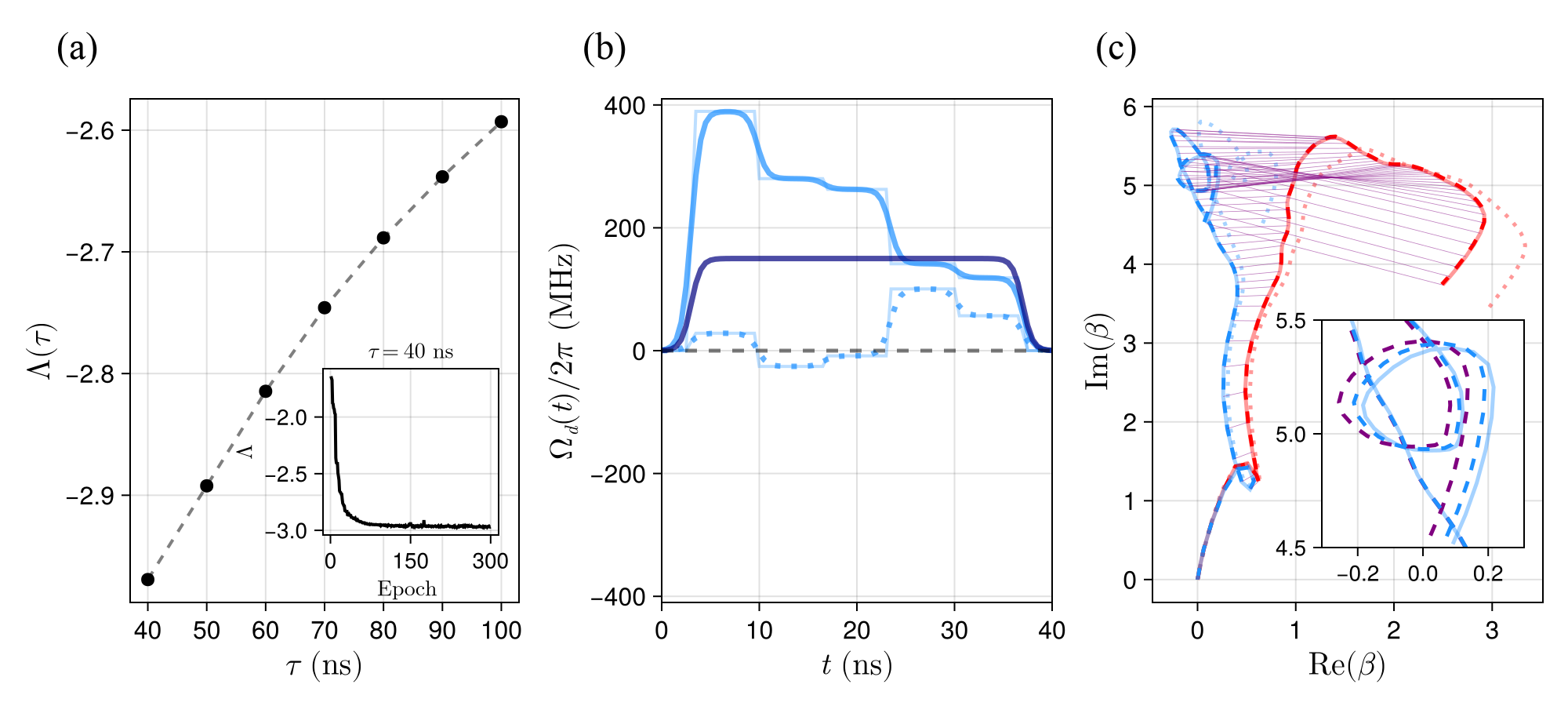}
    \caption{(a) The optimal value of the loss function for various readout times $\tau$ between $40$ and $100$ ns. Inset: a typical variation of the loss function along the optimization schedule for $\tau = 40$ ns. (b) Real (solid) and imaginary (dotted) parts of the optimal pulse envelope (light blue) for $\tau=40~\mathrm{ns}$. The envelope of the square pulse (dark blue) is plotted for comparison. (c) Phase-space trajectories for the optimal readout schedule and $\tau=40~\mathrm{ns}$, for initial states $\ket{g}$ (red) and $\ket{e}$ (blue), computed with $N_t = 5$, $N_u = 4$ and $N_l = 100$. The full and RWA trajectories are taken from a simulation in a Hilbert space of increased dimension $N_t = 6$, $N_u = 6$ and $N_l = 100$. Inset: difference between $M = 20$ (dashed purple) and $M = 80$ (dashed blue) low-rank evolutions compared to the full result (blue) in the region of phase-space where the discrepancy is most pronounced.}
    \label{fig:optimization}
\end{figure*}


\emph{Fidelity optimization.---}
Having established a method that accurately simulates a single instance of the transmon readout protocol at a minimal computational cost, we proceed by employing it to optimize the readout fidelity. This fidelity is mainly determined by the pulse shape and amplitude of the drive. In Ref.~\cite{gautier_optimal_2025}, the pulse envelope was represented by $1~\mathrm{ns}$ time bins, resulting in several hundreds of nominally independent parameters to optimize. We argue here that this is an unnecessarily complex parameterization scheme, as the time scale of the system response is considerably longer. In practice, pulse optimization often converges to physically meaningful waveforms that admit simple analytical descriptions with only a few free parameters~\cite{chatterjee_enhanced_2025}.

In what follows, we therefore parametrize the pulse as a stepwise complex-valued function with a smooth interpolation between its levels, as illustrated in Fig.~\ref{fig:schematic}b. We also assume that the pulse envelope $\Omega_d(t)$ is complex-valued, i.e. that the pulse can feature a slowly varying time-dependent phase. This amounts to a total of $2n_{\mathrm{steps}}$ parameters, where $n_{\mathrm{steps}}$ is the number of steps in the pulse. Here, we take $n_{\mathrm{steps}} = 5$, i.e. ten parameters to optimize. The validity of this approach is ultimately determined by the quality of the optimized readout protocol. We do not include the driving frequency $\omega_d$ as an optimization parameter, and set it at the eigenfrequency of the transmon–lower normal mode system~\cite{swiadek_enhancing_2024}.

Restricting the parameter space to a minimal number of physically relevant parameters brings several advantages. First, it allows for an easier and faster convergence of the optimization algorithm, in particular by enabling the use of computationally less demanding gradient-less optimization schemes, and by increasing the likelihood of finding a global, rather than a local minimum. Second, it opens the way for the optimization of more complex protocols where more intricate pulse schedules are required. Here, we found that the Simultaneous Perturbation Stochastic Approximation (SPSA)~\cite{spall_multivariate_1992, Bhatnagar_Prasad_Prashanth_2013}, leads to excellent optimization results, but we verified that similarly good results were obtained using a truly global optimization scheme such as the particle swarm algorithm~\cite{kennedy_pso_1995}. 

We determine the optimal readout schedule by searching for the drive pulse sequence that minimizes the loss function $\Lambda(\tau)$ for a given readout time $\tau$. During optimization, the loss was evaluated using the LRA with $M=20$. The RWA was not used during optimization. Fig.~\ref{fig:optimization}a shows the optimal value of the loss as a function of the assumed pulse duration $\tau$. The optimal value obtained for $\tau=40~\mathrm{ns}$ corresponds to an assignment error $\epsilon_a = 1.2\times 10^{-3}$, comparable with the current state of the art~\cite{swiadek_enhancing_2024, gautier_optimal_2025}. The inset displays a representative optimization schedule for $\tau=40~\mathrm{ns}$. We observe that the SPSA converges in $\mathcal{O}(100)$ epochs, corresponding to 200 evaluations of the loss, or 400 integrations of the master equation and a total wall-clock time $T_{\mathrm{LRA}}\simeq 1.2\times 10^{4}~\mathrm{s}$, i.e., less than 3 hours on the CPU of a modern laptop.

The optimal pulse envelope obtained for $\tau=40~\mathrm{ns}$ is shown in Fig.~\ref{fig:optimization}b. Notably, the real part of the optimized pulse resembles a two-step pulse: a strong initial drive that rapidly populates the readout cavity, followed by a reduced amplitude to limit undesired ionization events~\cite{walter_rapid_2017, mcclure_rapid_2016}. This result is obtained using only a single drive on the readout mode, without a secondary drive on the transmon~\cite{touzard_gated_2019, arias_qubit_2023, ikonen_qubit_2019}. Including these features as additional optimization parameters is expected to further reduce the optimal error.

The real and imaginary parts of the field in the filter resonator for the optimal pulse are shown in Fig.~\ref{fig:optimization}c. These were computed with the higher truncations $N_t = 6$, $N_u = 6$, and $N_l = 100$ in order to validate the optimized simulations. The full and LRA results essentially coincide except for a slight discrepancy at the end of the readout schedule for the trajectory initialized with the qubit in the excited state. The convergence of the LRA can be assessed by repeating the calculation for higher $M$. The inset of Fig.~\ref{fig:optimization}c shows the same calculation with $M=80$, which more closely approximates the exact result, as further discussed in Ref.~\cite{SM}. By contrast, the trajectories computed with the RWA differ substantially from the exact ones.

Restricting ourselves to controlled approximations---namely the Hilbert space truncations $N_t$, $N_u$, and $N_l$, and the rank $M$---allows the optimization to be run with a computationally faster yet more aggressive approximation. The rationale is that even if the loss is not quantitatively exact, it remains correlated with the value obtained under a more conservative approximation, so the resulting optimal parameters are expected to be reliable. To support this, in Ref.~\cite{SM} we repeat the full optimization with $M=80$ and find that they essentially coincide. By contrast, adopting a similar strategy with an uncontrolled approximation such as the RWA is more likely to yield inaccurate loss estimates and less reliable optima.


\emph{Conclusion.---}
In this letter, we have demonstrated that the low-rank approximation provides an accurate and fast approach to quantum optimal control of quantum device protocols. The essential reason is that, by construction, most of these protocols keep the system in high-purity states throughout their schedules. In this regime, a low-rank representation of the density matrix suffices to capture the dynamics with high accuracy. This enables a drastic reduction in computational time and memory, while preserving predictive power.

We have benchmarked this approach on the optimization of the transmon qubit readout protocol in a realistic transmon-resonator-filter model. Thanks to the LRA’s speed and a compact yet physically essential pulse parametrization, we recover a \emph{state-of-the-art assignment error} $\epsilon_a \approx 1.2\times10^{-3}$ at $40\,\mathrm{ns}$. This optimization could be carried out seamlessly on a standard laptop in two orders of magnitude less optimization time, without relying on uncontrolled approximations such as the RWA.

Beyond this specific application, the computational framework we introduced is broadly applicable. Any protocol whose dynamics remains weakly mixed---including high-fidelity gates and fast reset---can leverage its low rank to widen the reachable design space, lift ad-hoc approximations, and scale to larger Hilbert spaces and richer control schedules, offering a pathway to accelerate QOC across diverse quantum platforms. To facilitate its adoption, we provide an open-source implementation \cite{goutte_github_2025}, while a more general variational low-rank Lindblad solver is also available as part of QuantumToolbox.jl \cite{mercurio_quantumtoolboxjl_2025}. By making accurate, large-scale device optimization feasible with greatly reduced computational resources, this approach can dramatically broaden access to realistic QOC and accelerate progress toward scalable quantum technologies.


\section*{acknowledgments}
We acknowledge enlightening discussions with Lorenzo Fioroni, Filippo Ferrari, and Alberto Mercurio. We are grateful to Ronan Gautier for making the raw data of Ref.~\cite{gautier_optimal_2025} available to us. This work was supported by the Swiss National Science Foundation through Projects No. 200021-227992, 20QU-1\_215928, and 200020\_215172.




\appendix


\bibliographystyle{apsrev}
\bibliography{apssamp}

\include{supplemental_material}

\end{document}

%% file: supplemental_material.tex

\renewcommand{\thesection}{S\arabic{section}}
\renewcommand{\theequation}{S\arabic{equation}}
\renewcommand{\thefigure}{S\arabic{figure}}
\renewcommand{\thetable}{S\arabic{table}}

\setcounter{equation}{0}
\setcounter{figure}{0}
\setcounter{table}{0}
\setcounter{section}{0}





\clearpage

\onecolumngrid
\begin{center}
    \textbf{\large Supplemental Material for: ``Low-rank optimal control of quantum devices''}\\[0.5cm]
    Leo Goutte and Vincenzo Savona \\[0.2cm]
    \textit{Institute of Physics, École Polytechnique Fédérale de Lausanne (EPFL), CH-1015 Lausanne, Switzerland and}\\
    \textit{Center for Quantum Science and Engineering, École Polytechnique Fédérale}\\
    \textit{de Lausanne (EPFL), CH-1015 Lausanne, Switzerland}
\end{center}

\twocolumngrid


\section*{S1. LOW-RANK FORMALISM}\label{sec:low-rank}

The density matrix in a Hilbert space of dimension $N$ may always be written as a statistical ensemble of pure states
\begin{equation}\label{eq:rho_diagonal}
    \hat\rho = \sum_{j=1}^N q_j \ket{\psi_j}\bra{\psi_j},
\end{equation}
where $q_j$ are the probabilities (in descending order), with $\sum_{j=1}^Nq_j=1$, and $\ket{\psi_j}$ are the (normalized) states entering the statistical mixture. Note that states $\{\ket{\psi_j}\}$ form a complete set but are not necessary mutually orthogonal, and the decomposition Eq.~\eqref{eq:rho_diagonal} is not in general unique.

If the system under consideration is weakly mixed, the density matrix is well approximated by a truncated low-rank ansatz~\cite{donatella_continuous-time_2021, mccaul_fast_2021, chen_low-rank_2021, santos_low-rank_2025}:
\begin{equation}
\label{eq:low_rank_approx}
    \hat\rho \simeq \sum_{j=1}^M p_j \ket{\psi_j}\bra{\psi_j}\,,
\end{equation}
where $p_j={q_j}/{\sum_{k=1}^Mq_k}$ and $M \leq N$ is the rank of $\hat\rho$. The validity of this approximation is measured by the fidelity between states \eqref{eq:rho_diagonal} and \eqref{eq:low_rank_approx}, which should be close to unity along the system dynamics. Assuming that states $\ket{\psi_j}$ form an orthonormal basis, this translates to the minimization of the truncation error $\epsilon_M=1-\sum_{j=1}^Mq_j$.

\subsection{A. Low-rank equation of motion}
\label{sec:NOSSE}

The time evolution within the low-rank manifold is governed by the so-called non-stochastic matrix Schrodinger equation introduced in Refs.~\cite{joubert-doriol_non-stochastic_2014, joubert-doriol_problem-free_2015} which describes both the unitary and dissipative evolution of the states $\ket{\psi_j}$ in a fully deterministic manner. Here we briefly review the main formalism. 

We begin by writing the density matrix in the diagonal basis as
\begin{equation}
\label{eq:lowrank_diagonal}
    \hat\rho(t) = \mathbf m(t) \mathbf m^\dagger(t)
\end{equation}
where $\mathbf{m}(t) = (\sqrt{p_1(t)}\ket{\psi_1(t)}  \ \dots \ \sqrt{p_M(t)}\ket{\psi_M(t)})$ is a $N \times M$ matrix, $\ket{\psi_j(t)}$ are the (normalized) states entering the low-rank mixture, and $p_j(t)$ the corresponding probabilities, with $\sum_{j=1}^Mp_j=1$. Here and in the remainder of this work, we omit writing the explicit time-dependence unless ambiguous. The equation governing the dynamics of $\mathbf{m}$ can be derived starting from the master equation ($\hbar = 1$) 
\begin{subequations}\label{eq:master_equation}
\begin{align}
    \dot{\hat{\rho}} &= \LL[\hat\rho] =  -i \left[\hat H,\hat\rho\right] + \sum_k \mathcal{D}\left[\hat L_k\right] \hat\rho, \\
    \label{eq:lindblad_dissipator}
    \mathcal{D}[\hat L_k]\hat\rho &= \hat L_k \hat\rho \hat L_k^\dagger - \frac{1}{2}\left\{\hat L_k^\dagger \hat L_k, \hat\rho\right\},
\end{align}
\end{subequations}
where $\hat L_k$ are the Lindblad operators describing the dissipation processes. Substituting Eq.~\eqref{eq:lowrank_diagonal} into Eq.~\eqref{eq:master_equation} leads to an equation of motion for the matrix $\mathbf{m}$
\begin{equation}\label{eq:nosse}
    \dot{\mathbf{m}} = -i \hat H\mathbf{m} + \mathcal{O}[\mathbf{m}]\,.
\end{equation}
The first term on the right-hand side of Eq.~\eqref{eq:nosse} governs the unitary evolution of $\mathbf{m}$. The second term captures the dissipative non-unitary dynamics and is defined through the relation
\begin{equation}
\sum_k \mathcal{D}\left[\hat L_k\right] \mathbf{m}\mathbf{m}^\dagger = \mathcal{O}[\mathbf{m}] \mathbf{m}^\dagger + \mathbf{m} \mathcal{O}[\mathbf{m}]^\dagger.    
\end{equation}
Using Eq.~\eqref{eq:lindblad_dissipator}, we obtain the specific form of $\mathcal{O}[\mathbf{m}]$,
\begin{equation}
    \mathcal{O}[\mathbf{m}] = \frac{1}{2}\sum_k \hat L_k \mathbf{m} (\mathbf{m}^{-1} \hat L_k \mathbf{m})^\dagger - \hat L_k^\dagger \hat L_k \mathbf{m}.
\end{equation}
Note that $\mathbf{m}$ is a $N \times M$ rectangular matrix, whose inverse must be computed as the Moore-Penrose inverse defined as $\mathbf{m}^{-1} = (\mathbf{m}^\dagger \mathbf{m})^{-1} \mathbf{m}^\dagger$. The term Low-Rank Approximation (LRA) in the main text refers to both Eq.~\eqref{eq:low_rank_approx} and Eq.~\eqref{eq:nosse}. 

Here, we assume the rank $M$ to be constant in time. The evolution of our system according to Eq.~\eqref{eq:nosse} therefore requires the multiplication of extremely sparse $N \times N$ matrices (such as $\hat H$ and $\hat L_k$) and dense $N \times M$ matrices (such as $\mathbf{m}$) at every time step. Conversely, a full evolution according to the master equation includes, once vectorized~\cite{manzano_lindblad_2020}, dense vectors of size $N^2$, and sparse matrices of size $N^2 \times N^2$. If $M \ll N$, both the compute time and memory requirements are therefore drastically improved. 

When assuming a fixed rank, care must be taken in setting the initial condition for the numerical integration of Eq.~\eqref{eq:nosse}. If the initial state has a rank $M^\prime<M$, as is the case here when choosing an initial pure state, then the matrix $\mathbf{m}^\dagger \mathbf{m}$ is singular and the Moore-Penrose inverse of $\mathbf{m}$ is not defined. We solve this issue by setting the states $\ket{\psi_j(0)}$, for $j=2,\ldots,\,M$ to a set of random orthonormalized states and the corresponding probabilities to small values $p_j=\epsilon\ll1$. The validity of this approach can be easily verified by checking the convergence of the results for decreasing values of $\epsilon$. For the study that follows, we find that $\epsilon=10^{-5}$ always guarantees convergence.

\subsection{B. Validity}\label{sec:validity}

Ideally, the validity of the LRA is measured by the truncation error $\epsilon_M$. Computing the low-rank density matrix \eqref{eq:lowrank_diagonal} by solving Eq.~\eqref{eq:nosse} however doesn't provide access to the the probabilities $q_j$ and therefore to $\epsilon_M$. Several control quantities can be used as proxies for the truncation error \cite{gravina_adaptive_2024}. Here, as a control quantity, we adopt the ratio $p_M / p_1$, which is readily computed as a function of time and closely reflects the accuracy of the LRA \cite{gravina_adaptive_2024}. We further keep track the purity $\mathrm{tr}(\hat\rho^2)$, which is overestimated by the LRA and is expected to converge to the exact value as the rank $M$ is increased. 

To illustrate the validity of the LRA and the reliability of the control quantities, we simulate a typical schedule of the dispersive readout of a transmon qubit, as described in detail in the main text. We compare the values of $p_M / p_1$ and of the purity $\mathrm{tr}(\hat\rho^2)$, as obtained from the full integration of the Lindblad master equation and from the LRA. 

We additionally integrate the Lindblad master equation without applying the LRA but instead applying the RWA, which consists in neglecting all counter-rotating terms, both in the coupling and drive terms, in the frame rotating with the driving field~\cite{blais_circuit_2021, khezri_measurement-induced_2023} (a full derivation is included in Sec.~\ref{sec:hamiltonian_rwa}). The RWA is often adopted in QOC as it leads to a large speedup in computational time due to the elimination of rapidly oscillating terms in the equation of motion~\cite{gautier_optimal_2025}. It is, however, known to produce quantitatively inaccurate results in cases when modes are nonlinear, when driving fields are not weak, or are off-resonant~\cite{kofman_unified_2004,shillito_dynamics_2022, dumas_measurement-induced_2024, sank_measurement-induced_2016}. It may therefore overlook phenomena which occur on fast timescales such as chaotic dynamics and transmon ionization~\cite{ferrari_dissipative_2025, cohen_reminiscence_2022, shillito_dynamics_2022, dumas_measurement-induced_2024}, and could lead to quantitatively inaccurate predictions already for moderate driving field amplitudes. It is also an uncontrolled approximation, as it can't be tuned by a control parameter, differently from LRA where instead simulations can be repeated for different values of the control parameter $M$, until convergence of the control quantities.

\begin{figure}
    \centering
    \includegraphics[width=0.8\linewidth]{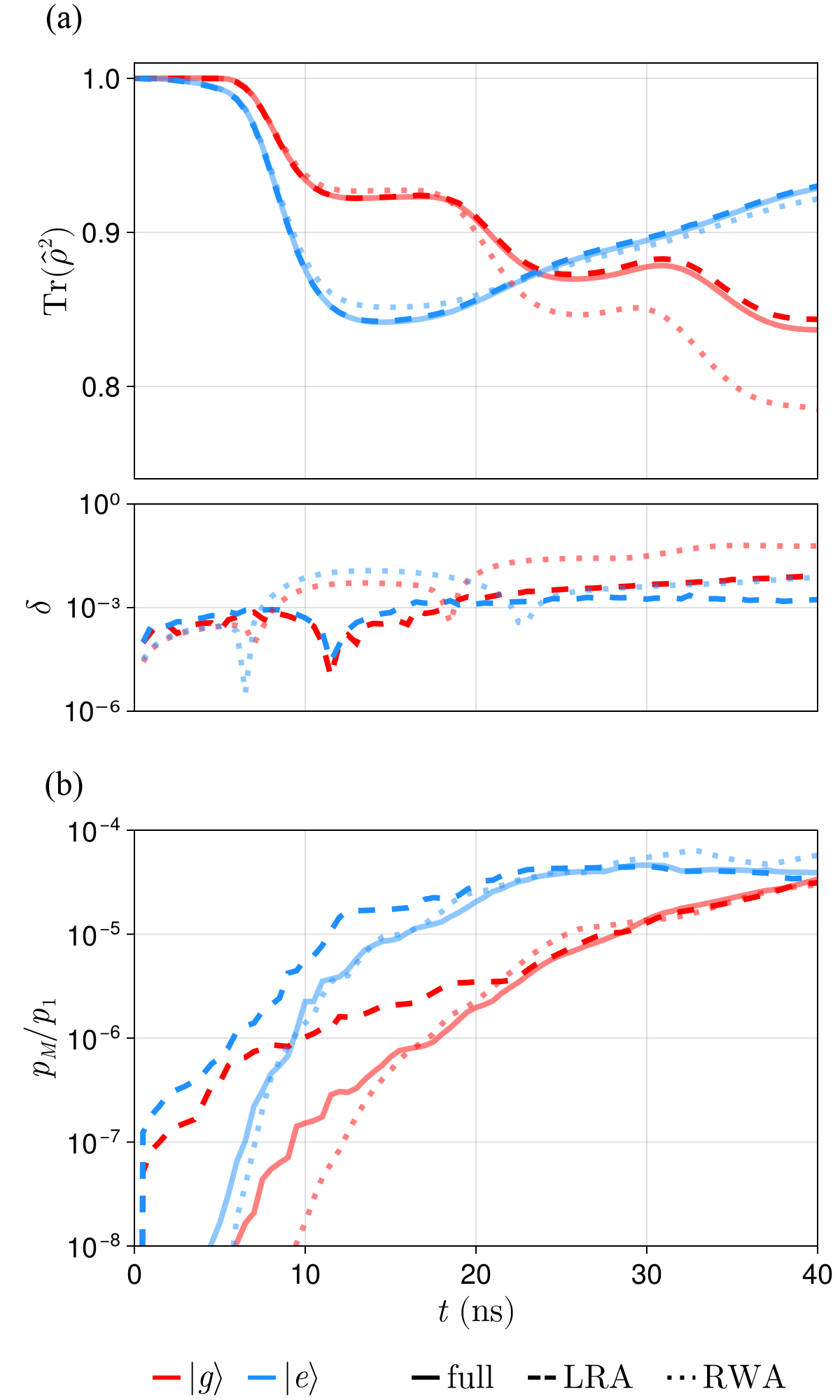}
    \caption{Time evolution of (a) the purity $\mathrm{tr}(\hat\rho^2)$ and (b) the control quantity $p_M/p_1$, as computed from integrating the full master equation Eq.~\eqref{eq:hamiltonian}, the LRA Eq.~\eqref{eq:nosse} with $M = 20$, and the RWA Eq.\eqref{eq:hamiltonian_rwa} for the transmon readout. The solid, dashed, and dotted lines correspond to the full evolution, low-rank, and rotating-wave approximation evolutions, respectively. The red (blue) lines correspond to an initial transmon ground (first excited) state. The purity generally decreases in time as the system evolves towards a mixed state. The relative error between the full and approximate values $\delta$ is much smaller for the LRA than for the RWA. (b) The control quantity $p_M/p_1$ increases as the state becomes mixed, providing a practical indicator of the LRA’s accuracy.}
    \label{fig:purity}
\end{figure}

Fig.~\ref{fig:purity} shows the results obtained by integrating the equations of motion for three levels of approximation: the full Lindblad master equation, the LRA, and the RWA. The dynamics are simulated in presence of a square driving pulse of duration $40$ ns, in the same setting as illustrated in Fig.~\ref{fig:schematic}. For this simulation, a Hilbert space of total dimension $N=2000$ was considered. For the LRA, $M=20$ was taken. This choice of parameters coincides with that leading to the results presented in Fig.~\ref{fig:phase_space_trajectories}. The comparison presented in Fig.~\ref{fig:purity} shows how the LRA essentially coincides with the full simulation already for such a small value of $M$. The RWA on the other hand leads to a sizable quantitative inaccuracy. The LRA can thus essentially reproduce the exact results, and hence be predictive for QOC applications, provided the convergence of the results versus $M$ is carefully checked, or alternatively an adaptive low-rank scheme is used~\cite{gravina_adaptive_2024}.

\section{S2. DETAILS OF THE SIMULATION}\label{sec:sim_details}


In what follows, we provide some additional details to facilitate the reproduction of the results in the main text.

\subsection{A. Hilbert space}\label{sec:hilbert_space}

To build the Hilbert space basis on which Eq.~\eqref{eq:hamiltonianpurcell} is numerically expressed, we first diagonalize the transmon Hamiltonian $\hat H_t = 4E_c \hat n_t^2 - E_j \cos\hat\varphi_t$ in the flux basis, i.e. by assuming the canonical commutation relation $[\hat \varphi_t, \hat n_t] = i$. The transmon Hamiltonian is then rewritten as $\hat H_t = \sum_{j_t=0}^{N_t-1} \omega_{j_t} \ket{j_t}\bra{j_t}$, where $\ket{j_t}$ is the $j_t$th transmon eigenstate and only the lowest $N_t$ eigenstates are retained. 
We also diagonalize the resonator-filter subsystem in the normal-mode basis $\{\ket{j_l}\otimes\ket{j_u}\}$, where $\ket{j_l}$ and $\ket{j_u}$ are the eigenstates of the lower and upper normal modes, respectively, with frequencies $\omega_l$ and $\omega_u$. For readout we choose the lower normal mode. Correspondingly, the Fock eigenspace of the lower normal mode is truncated to a large number of Fock states, up to $N_l = 100$, in order to account for its potentially large photon number. The other modes are off-resonant and are not expected to reach a large occupation. We therefore opt for smaller cutoffs in the transmon and upper normal modes, up to $N_t = 6$ and $N_u = 6$, respectively. The convergence of all the results has been checked with respect to these cutoffs.

We finally build the full Hilbert space of the coupled system as the tensor product of the transmon and normal-mode eigenstates, i.e. $\{\ket{j_l}\otimes\ket{j_u}\otimes\ket{j_t}\}$, where $j_l = 0, \ldots, N_l - 1$, $j_u = 0, \ldots, N_u - 1$, and $j_t = 0, \ldots, N_t - 1$, and a total Hilbert space dimension $N=N_l\times N_u\times N_t$.

For the readout procedure, the state is initialized in the vacuum of the normal modes and the ground/first excited state of the transmon,
\begin{align}\label{eq:initial_states}
    \hat\rho_{g/e}(0) &= \ket{\psi_{g/e}}\bra{\psi_{g/e}},\\
    \ket{\psi_{g/e}} &= \ket{0_l} \otimes \ket{0_u} \otimes \ket{g/e}\,.
\end{align}
Eq.\eqref{eq:nosse} is then numerically integrated for a total readout time $\tau$ and expectation values of fields and occupations are evaluated. 

\subsection{B. Parameters}\label{sec:parameters}

Throughout this work and unless otherwise stated, we set the transmon parameters to $E_c/2\pi = 315$ MHz and $E_j/E_c = 51$. This yields a qubit frequency $\omega_t /2\pi = 6$ GHz and anharmonicity $\alpha / 2\pi = -349$ MHz. We take the couplings $g/2\pi = 150$ MHz and $J/2\pi = 30$ MHz. The frequency of the resonator is $\omega_r/2\pi = 7.2$ GHz while that of the filter is $\omega_f/2\pi = 7.21$ GHz. The dissipation rates are $\kappa/2\pi = 30$ MHz and $\gamma/2\pi = 8$ kHz, for the transmon and filter, respectively. The frequency of the drive is taken to be $\omega_d / 2\pi = 7.18$ GHz.

\subsection{C. Drive pulse shape}
\label{sec:drive_pulse}

As mentioned in the main text, a microwave drive pulse is applied to the Purcell filter modes at a frequency $\omega_d$. The drive pulse may generally have a time-dependent, complex envelope given by $\Omega_d(t)$. In this work, we begin by setting $\Omega_d(t)$ to a square pulse shape and later generalize it to contain many steps.

For the initial simulations with a square pulse shape, $\Omega_d(t)$ is given by
\begin{equation}
\label{eq:square_pulse}
    \Omega_d(t) = |\Omega_d| \times f(t,t_0,\sigma) \times \left[1 - f(t,t_0 + T,\sigma)\right]
\end{equation}
where
\begin{equation}
    f(t, t_0, \sigma) = \frac{1}{1 + e^{-\frac{t-t_0}{\sigma}}}
\end{equation}
is the logistic function~\cite{arfken_mathematical_2013}. The parameters $t_0$ and $\sigma$ set the curve's center and steepness, respectively. The time it takes to go from an amplitude of zero to $|\Omega_d|$, referred to in the main text as ramp up time, is given by $\sim 5 \sigma$. In the present work, we choose $t_0 = 3$ ns and $\sigma = 0.5$ ns, corresponding to a ramp up/down time of $2.5$ ns. 

The square pulse is then straightforwardly generalized to a smoothed stepwise function. For a given list of $n_{\mathrm{steps}}$ heights $h$ and at a time $t$, the smoothed stepwise function outputs $ \sum_{j = 1}^{n_{\mathrm{steps}}}(h_{j+1} - h_{j}) \times f(t, j\tau/n_{\mathrm{steps}},\sigma)$. An illustrative example of both the square and smoothed stepwise functions is shown in Fig.~\ref{fig:schematic}b.


\subsection{D. Diagonalization of $\hat H_{\mathrm{rf}}$}\label{sec:normal_modes}

We seek two normal modes $\hat c_l$ and $\hat c_u$ such that
\begin{equation}
    \hat H_{\mathrm{rf}} = \omega_r \hat a^\dagger \hat a + \omega_f \hat f^\dagger \hat f - J(\hat a^\dagger - \hat a)(\hat f^\dagger - \hat f)
\end{equation}
can be written as
\begin{equation}
   \hat H_{\mathrm{rf}} = \omega_l \hat c_l^\dagger \hat c_l + \omega_u \hat c_u^\dagger \hat c_u.
\end{equation}
Therefore, we manifestly have
\begin{equation}
\label{eq:normalmodecommutation}
    [\hat H_{\mathrm{rf}}, \hat c_i] = -\omega_i \hat c_i.
\end{equation}
We define the normal modes in terms of the bare operators $\hat a$ and $\hat f$ as follows:
\begin{equation}
\label{eq:normalmodedef}
    \hat c_i = \alpha_i \hat a + \beta_i \hat f + \gamma_i \hat a^\dagger + \delta_i \hat f^\dagger.
\end{equation}
Note that we must account for the particle-hole nature of the $c_i$ modes due to the presence the terms $-J \hat a^\dagger \hat f^\dagger - J \hat a \hat f$. This in turn yields the symmetry $\hat c_i \to \hat c_i^\dagger$ bringing $\omega \to -\omega$ following directly from Eq.~\eqref{eq:normalmodecommutation}. 

Substituting Eq.~\eqref{eq:normalmodedef} into Eq.~\eqref{eq:normalmodecommutation}, we find four equations for the four coefficients of $\hat c_i$ by imposing that the terms attached to each operator $\hat a$, $\hat f$, $\hat a^\dagger$ and $\hat f^\dagger$ must match. These equations can succinctly written in matrix form
\begin{equation}
    \mathbf\Omega = \begin{pmatrix}
        \omega_r & J & 0 & J \\
        J & \omega_f & J & 0 \\
        0 & -J & -\omega_r & -J \\
        -J & 0 & -J & -\omega_f
    \end{pmatrix}.
\end{equation}
Diagonalizing $\mathbf\Omega$ yields the two normal modes $\omega_l$ and $\omega_u$ along with their respective negative counterparts due to the aforementioned symmetry.

We now seek the transformation from $\hat a$, $\hat f$ to $\hat c_l$, $\hat c_u$ in order to express our Hamiltonian in terms of the normal modes. To do so, we write the system of equations
\begin{equation}
    \mathbf{c} = U^\top \mathbf{a}
\end{equation}
which express the vector of operators $\mathbf{c} = (\hat c_u^\dagger, \hat c_l^\dagger, \hat c_l, \hat c_u)^\top$ as linear combinations of $\mathbf{a} = (\hat a,\hat f,\hat a^\dagger,\hat f^\dagger)^\top$. Note the ordering of these operators is done to match the ordered list of frequencies, from $-\omega_u$ to $\omega_u$, with $\omega_u > \omega_l$. The matrix $U$ contains in each of its columns the eigenvectors of $\mathbf\Omega$ normalized subject to the condition
\begin{equation}
    |\alpha|^2 + |\beta|^2 - |\gamma|^2 - |\delta|^2 = 1,
\end{equation}
given by the commutation relations of the normal operators $[c_i,c_j^\dagger] = \delta_{ij}$. Note the difference with respect to the naive normalization $|\alpha|^2 + |\beta|^2 + |\gamma|^2 + |\delta|^2 = 1$. One can verify that the commutation relations are preserved by this transformation via the equality
\begin{equation}
    S \mathbf{\Sigma}_c S^\dagger = {\mathbf{\Sigma}}_a
\end{equation}
where we have defined $S = \mathrm{inv} \ U^\top$ such that $\mathbf{a} = S\mathbf{c}$, and 
\begin{align}
\begin{split}
\label{eq:symplecticmatrices}
    \left(\mathbf{\Sigma}_a\right)_{ij} &= \left[a_i, a_j\right] \Rightarrow\begin{pmatrix}
        0 & \mathbf{1} \\
        -\mathbf{1} & 0
    \end{pmatrix}  , \\
    \left(\mathbf{\Sigma}_c\right)_{ij} &= \left[c_i, c_j\right] \Rightarrow \begin{pmatrix}
        0 & -\sigma_x \\
        \sigma_x & 0
    \end{pmatrix}.
\end{split}
\end{align}

The coefficients of the transformation from the bare to the normal operators are the elements of the matrix $S$. Letting $(S)_{ij} = s_{ij}$, we can in particular denote
\begin{align}
    \hat a &= s_{11} \hat c_u^\dagger + s_{12} \hat c_l^\dagger + s_{13} \hat c_l + s_{14} \hat c_u, \\
    \hat f &= s_{21} \hat c_u^\dagger + s_{22} \hat c_l^\dagger + s_{23} \hat c_l + s_{24} \hat c_u,
\end{align}
and so on for $a^\dagger$ and $f^\dagger$. Without loss of generality, we take $s_{ij}$ to be fully real. The quantities of interest in the coupling and drive, $a^\dagger - a$ and $f^\dagger - f$, respectively, take the following forms:
\begin{align}
    a^\dagger - a &= (s_{14} - s_{11}) \left(\hat c_u^\dagger - 
    \hat c_u \right) + (s_{13} - s_{12}) \left(\hat c_l^\dagger - 
    \hat c_l \right), \\
    f^\dagger - f &= (s_{24} - s_{21}) \left(\hat c_u^\dagger - 
    \hat c_u \right) + (s_{23} - s_{22}) \left(\hat c_l^\dagger - 
    \hat c_l \right).
\end{align}
We can finally define the constants $\mu_u = s_{14} - s_{11}$, $\mu_l = s_{13} - s_{12}$, $\nu_u = s_{24} - s_{21}$, and $\nu_l = s_{23} - s_{22}$.


\subsection{E. Rotating-Wave Approximation on $\hat H$}
\label{sec:hamiltonian_rwa}

We seek to isolate and eliminate the fast rotating terms from our Hamiltonian Eq.~\eqref{eq:hamiltonianpurcell},
\begin{equation}\label{eq:hamiltonian2}
    \hat H = \hat H_t + \hat H_{\mathrm{rf}} +ig\hat n_t \left(\hat a^\dagger - \hat a\right) + i \Omega_d(t) \sin(\omega_d t)\left(\hat f^\dagger - \hat f\right).
\end{equation}
To identify these terms, we begin by expressing $\hat H$ in terms of the normal modes $\hat c_l$ and $\hat c_u$: 
\begin{equation}
    \begin{split}
    \hat H =& \hat H_t + \hat H_{\mathrm{rf}} + ig \hat n_t \left[\mu_u\left(\hat c_u^\dagger - \hat c_u\right) + \mu_l\left(\hat c_l^\dagger - \hat c_l\right)\right] \\
    &+ i\Omega_d(t)\sin(\omega_d t) \left[\nu_u\left(\hat c_u^\dagger - \hat c_u\right) + \nu_l\left(\hat c_l^\dagger - \hat c_l\right)\right]
\end{split}
\end{equation}
We then rotate into the frame of the drive by use of the unitary
\begin{equation}
   \hat U_d =\hat U_l \otimes \hat U_u \otimes \hat U_t
\end{equation}
where $\hat U_{l/u} = \exp(i \omega_d t \hat c^\dagger_{l/u}\hat c_{l/u})$ and 
\begin{equation}
    \hat U_t = \exp[i\omega_d t \sum_{j=1}^{N_t} (j-1)\ket{\psi_j}\bra{\psi_j} + i\omega_1 t]
\end{equation}
and $\omega_1$ is the ground state energy of the transmon Hamiltonian $H_t$. The Hamiltonian transforms according to $\hat H \to \hat U_d^\dagger \hat H \hat U_d - i \hat U_d^\dagger \dot{\hat{U_d}}$. 

The normal mode operators will simply acquire a time-dependent phase: $\hat c_{l/u} \to \hat c_{l/u} e^{-i\omega_d t}$. To elucidate the transformation of $\hat n_t$, it is useful to decompose it in terms of its counter-clockwise- and clockwise-rotating terms
\begin{equation}
    \hat n_t = \sum_{j>k} \hat n_t^{jk} \ket{j}\bra{k} + \sum_{j<k} \hat n_t^{jk} \ket{j}\bra{k} = \hat n_t^- + \hat n_t^+ .
\end{equation}
The case $j=k$ is omitted since $\hat n_t$ contains only zeros on its diagonal. We then make the approximation $\hat n_t^+ \simeq \sum_{j} \hat n_t^{j,j+1} \ket{j}\bra{j+1}$ and similarly for $\hat n_t^-$. In doing so, $\hat n_t^+$ now transforms in the familiar way $\hat n_t^+ \to \hat n_t^+ e^{-i\omega_d t}$. 

The total Hamiltonian now transforms in a straightforward way. First, the bare Hamiltonians become
\begin{align}
    \hat H_t^\prime &=  \sum_{j = 1}^{N_t} \left[\omega_j - \omega_1 - \omega_d (j - 1)\right] \ket{j}\bra{j}, \\
     \hat H_{\mathrm{rf}}^\prime &= \left(\omega_l - \omega_d\right) \hat c_l^\dagger \hat c_l + \left(\omega_u - \omega_d\right) \hat c_u^\dagger \hat c_u.
\end{align}
The Hamiltonian in this rotated frame is then
\begin{equation}\label{eq:hamiltonian_rotated}
\begin{split}
        \hat H_{\mathrm{rot}} =& \hat H_t^\prime + \hat H_{\mathrm{rf}}^\prime \\
    &+ ig\mu_u \left(\hat n_t^+ \hat c_u^\dagger - \hat n_t^- \hat c_u - \hat n_t^+ \hat c_u e^{-2i\omega_d t} + \hat n_t^- \hat c_u^\dagger e^{2i\omega_d t} \right)  \\
    &- \frac{\nu_u}{2} \Omega_d(t)  \left(\hat c_u^\dagger + \hat c_u - \hat c_u^\dagger e^{2i\omega_d t} - \hat c_u e^{-2i\omega_d t}\right) \\
    &+ (u \leftrightarrow l),
\end{split}
\end{equation}
where the last line should be understood as repeating the second and third lines but with the index $u$ replaced by $l$. Eq.~\eqref{eq:hamiltonian_rotated} is the Hamiltonian we use for the full and LRA simulations in the main text.

To obtain the RWA Hamiltonian, we simply ignore all explicitly time-dependent terms in the coupling and drive and arrive at the slow Hamiltonian
\begin{align}
    \label{eq:hamiltonian_rwa}
    \begin{split}
        \hat H_{\mathrm{rwa}} =& \hat H_t^\prime + \hat H_{\mathrm{rf}}^\prime + i g \mu_{l} \left(\hat n_t^+ \hat c_l^\dagger - \hat n_t^- \hat c_l  \right) \\
        & - \frac{\nu_{l}}{2}\Omega_d(t) \left(\hat c_l + \hat c_l^\dagger\right) + i g \mu_{u} \left(\hat n_t^+ \hat c_u^\dagger - \hat n_t^- \hat c_u \right) \\
        &  - \frac{\nu_u}{2}\Omega_d(t) \left(\hat c_u + \hat c_u^\dagger \right).
    \end{split}
\end{align}

We draw particular attention to the factor of $1/2$ appearing in the driving terms proportional to $\Omega_d(t)$, originating from the sine function in the Hamiltonian~\eqref{eq:hamiltonian2}. This departs from the usual convention of writing a RWA driving Hamiltonian as $\hat H_d^{\mathrm{rwa}}(t) = \Omega_d^{\mathrm{rwa}} (t) (\hat f + \hat f^\dagger)$. In our convention therefore, values of $\Omega_d(t)$ are twice as large as those quoted in Ref.~\cite{gautier_optimal_2025}. 

\section{S3. FIDELITY OPTIMIZATION DETAILS}\label{sec:optimization_details}

\subsection{A. Assignment error}\label{sec:assignment_error}

\begin{figure}[ht]
    \centering
    \includegraphics[width=\linewidth]{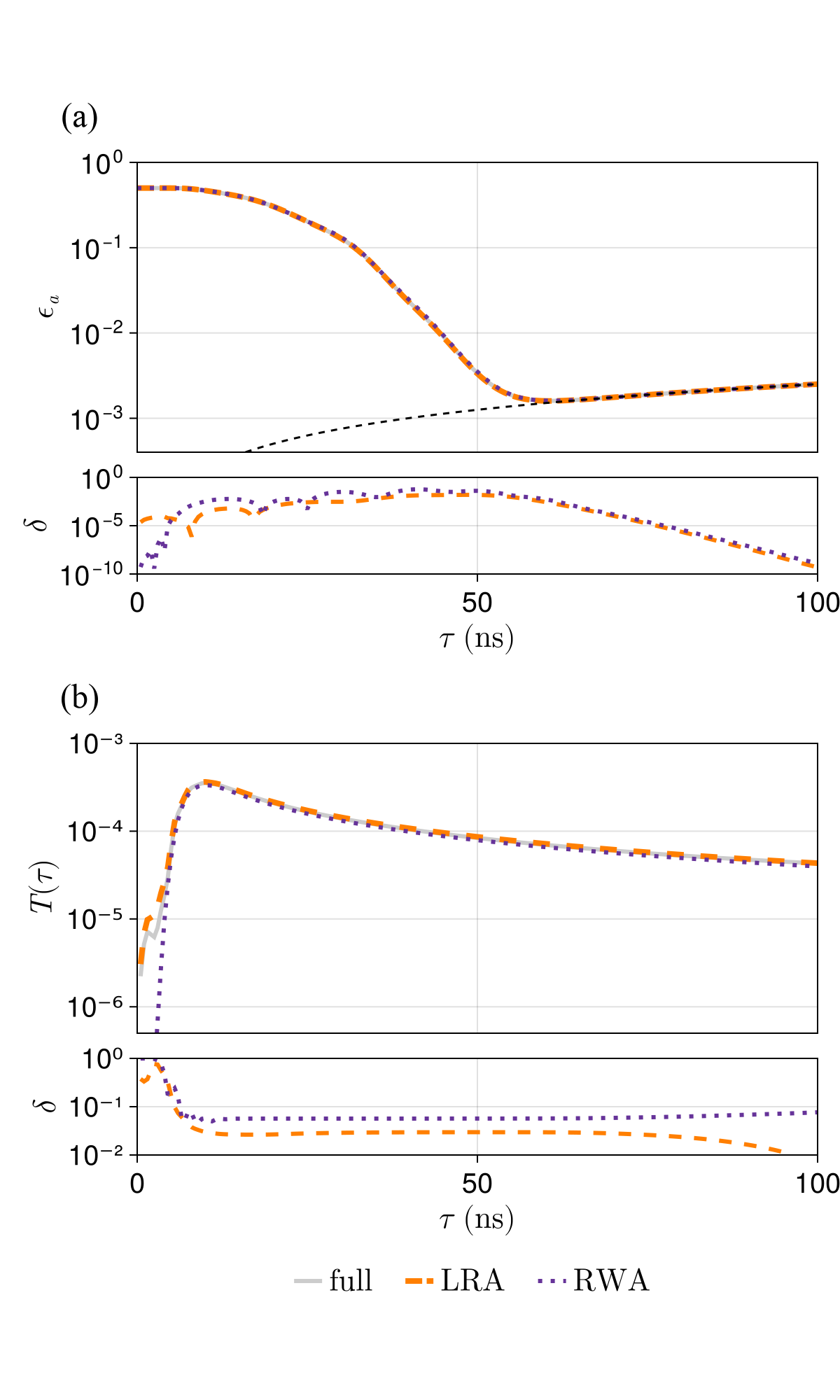}
    \caption{Transmon readout error and ionization for a square pulse compared to the LRA with rank $M=20$ and RWA. The full, low-rank, and RWA evolutions are plotted in grey, dashed orange, and dotted turquoise lines, respectively. (a) The assignment error $\epsilon_a(\tau)$ as a function of the readout time. The dashed black line corresponds to the long-$\tau$ limit $\epsilon_a \simeq  \tau\gamma/2$. The relative error $\delta$, shown in the lower panel, demonstrates that the LRA captures the exact assignment error slightly better than the RWA, though the RWA still performs well due to the relatively weak drive strength. (b) The transmon ionization Eq.~\eqref{eq:transmon_ionization}. Again, the LRA is a better match to the full dynamics than the RWA, though the latter still manages to capture the transmon ionizations since the drive is relatively weak.}
    \label{fig:snr}
\end{figure}

The trajectories of the readout mode in phase space are directly related to the fidelity of the readout measurement. The assignment error of the measurement is determined by the separation between the pointer states of the readout mode~\cite{gambetta_protocols_2007, bengtsson_model-based_2024}, and is expressed as  
\begin{equation}\label{eq:separation_error}
    \epsilon_{\mathrm{sep}}(\tau) = \frac{1}{2}\mathrm{erfc}\left[\frac{\mathrm{SNR}(\tau)}{2}\right].
\end{equation}
Here, the signal-to-noise ratio is defined as
\begin{equation}
\label{eq:SNR}
    \mathrm{SNR}(\tau) = \sqrt{2\eta\kappa \int_0^\tau dt \left|\beta_e(t) - \beta_g(t)\right|^2},
\end{equation}
where $\eta$ is the measurement efficiency which we set to be $\eta = 0.6$ to match with experiment \cite{walter_rapid_2017}, and $\beta_{e,g}(t)$ are the expectation values of the field in the readout mode, which are measured in the experiment. Together with the decay of the transmon from its excited to ground state captured by $\epsilon_{\mathrm{decay}} (\tau) = \tau \gamma / 2$, we obtain the total assignment error
\begin{equation}\label{eq:assignment_error}
    \epsilon_a(\tau) = \epsilon_{\mathrm{sep}}(\tau) + \epsilon_{\mathrm{decay}}(\tau)
\end{equation}
which is related to the readout fidelity by $\mathcal{F} = 1 - \epsilon_a$~\cite{bultink_general_2018, bengtsson_model-based_2024}. The competition between $\epsilon_{\mathrm{decay}}$, which favors short times, and $\epsilon_{\mathrm{sep}}$, which favors strong drives and long times, makes for an optimal readout time at which the assignment error is minimized, as seen in Fig.~\ref{fig:snr}a. For the square pulse, we find a minimum assignment error of $\epsilon_a = 1.6 \times 10^{-3} $ for a readout time of $60$ ns. A similar result is obtained with both the LRA and RWA.

\subsection{B. Loss function}\label{sec:loss_function}
We define a loss function that balances two competing effects: minimizing the assignment error while avoiding transmon ionization. The ionization penalty is quantified by the population in higher transmon levels~\cite{gautier_optimal_2025}:
\begin{equation}\label{eq:transmon_ionization}
    T (\tau) =  \frac{1}{\tau} \sum_{j = g, e} \sum_{k \geq 2}^{N_t} \int_0^{\tau}dt \ \mathrm{tr}\left[\ket{k_t}\bra{k_t} \hat\rho_j(t)\right].
\end{equation}
Here, $\hat\rho_j(t)$ is the time-evolved density matrix of the full system, obtained for an initial transmon state $\ket{j=g,e}$, respectively. The transmon ionization during readout with a square pulse is shown in Fig.~\ref{fig:snr}b. 
The loss function to minimize is finally
\begin{equation}
\label{eq:loss_function}
    \Lambda (\tau) = \log_{10}\left[\epsilon_a(\tau) + T(\tau)\right]\,,
\end{equation}
where the logarithm is taken to ensure that the loss function is of order one. 

We restrict the loss function to only these two contributions, as they are the most relevant especially for increasing drive strength. A more complete multi-objective optimization may include additional contributions to penalize populating the undriven normal mode or approaching the critical photon number~\cite{gautier_optimal_2025}. We find that the contribution of these terms to the loss function is minimal for the case studied here, and we therefore expect the minima of Eq.~\eqref{eq:loss_function} to remain close to those of a more comprehensive multi-objective loss function. Nevertheless, in order to safeguard against any further unaccounted ionization in the transmon and undriven normal mode, we impose an upper bound on the driving field strength as $|\Omega_d^{\mathrm{max}}|/2\pi < 400$ MHz. This value coincides with that used in Ref.~\cite{gautier_optimal_2025}. We also verify that the optimal result is characterized by very small values of these additional contributions.

\subsection{C. Optimization process}
\label{sec:optimization_process}

\begin{figure*}
    \centering
    \includegraphics[width=0.9\linewidth]{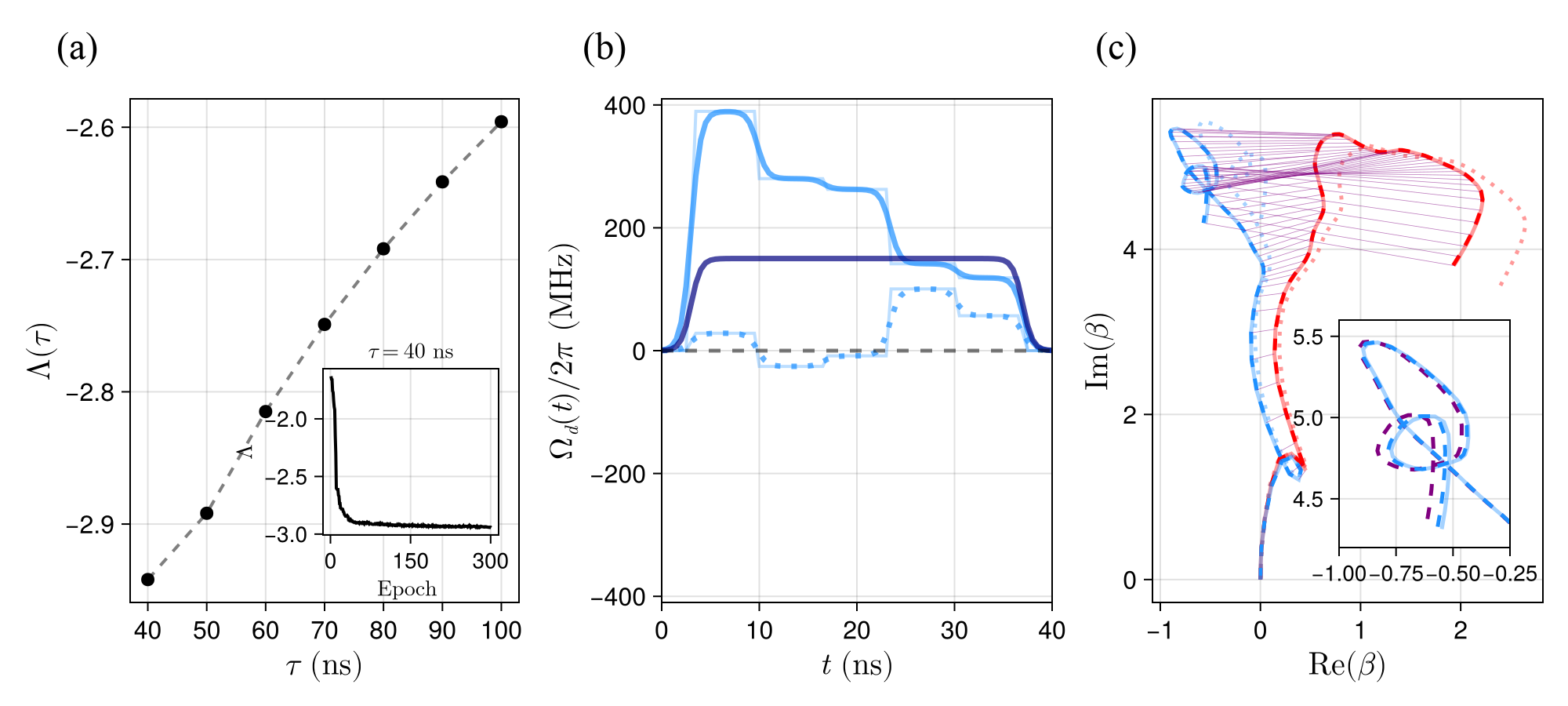}
    \caption{Optimization of the measurement fidelity performed with a rank $M = 80$. (a) The optimal value of the loss function for various readout times $\tau$ between $40$ and $100$ ns. Inset: a typical variation of the loss function along the optimization schedule for $\tau = 40$ ns. A minimum is reached in roughly $100$ epochs. (b) Real (solid) and imaginary (dotted) parts of the optimal pulse envelope (light blue) for $\tau=40~\mathrm{ns}$. The envelope of the square pulse (dark blue) is plotted for comparison. Thin lines show the stepwise pulse before smoothing. (c) Phase-space trajectories for the optimal readout schedule and $\tau=40~\mathrm{ns}$, computed with $N_t = 5$, $N_u = 4$ and $N_l = 100$. The full and RWA trajectories are taken from a simulation in a Hilbert space of increased dimension $N_t = 6$, $N_u = 6$ and $N_l = 100$. Inset: difference between $M = 20$ (dashed purple) and $M = 80$ (dashed blue) low-rank evolutions compared to the full result (blue) in the region of phase-space where the discrepancy is most pronounced.}
    \label{fig:optimization_80}
\end{figure*}

\begin{figure}[ht]
    \centering
    \includegraphics[width=0.9\linewidth]{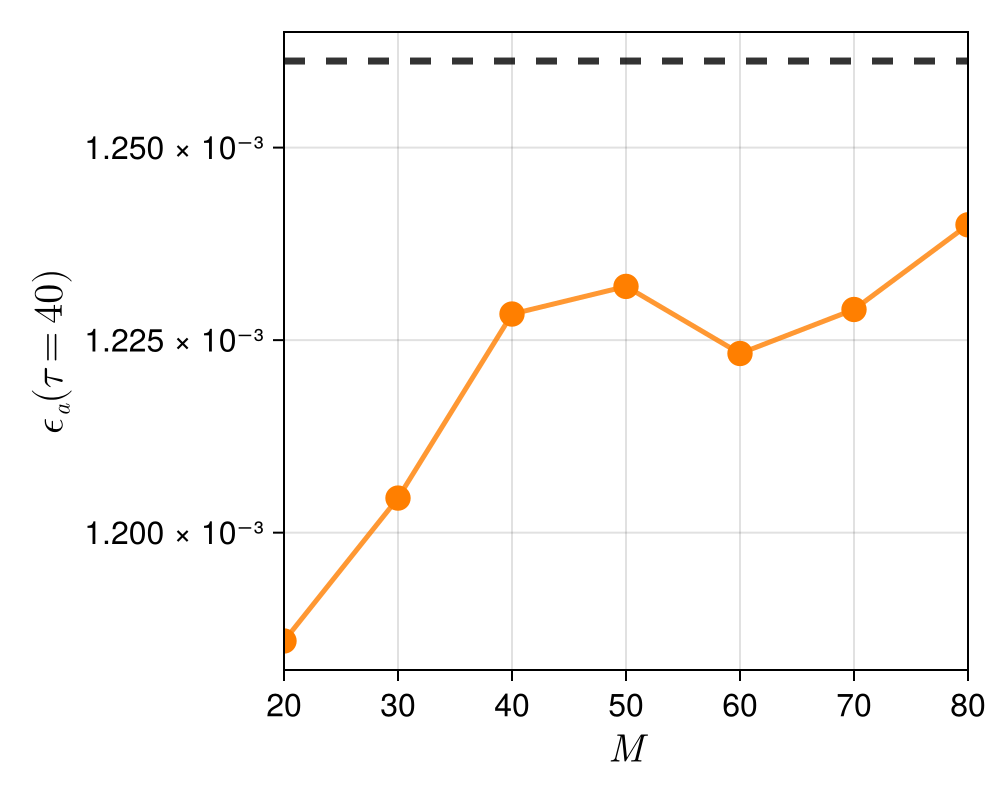}
    \caption{The assignment error Eq.~\eqref{eq:assignment_error} computed for $\tau = 40$ ns and varying rank $M$. The dashed black line at $\epsilon_{a} = 1.26 \times 10^{-3}$ corresponds to the assignment error obtained by integrating the full master equation.}
    \label{fig:SNR_Ms}
\end{figure}

We optimize the measurement fidelity by minimizing the loss function Eq.~\eqref{eq:loss_function} for a given readout time $\tau$. The optimization parameters are the drive pulse heights, which generally consist of both real and imaginary parts. for $n_{\mathrm{steps}}$ steps, there are therefore $2 n_{\mathrm{steps}}$ parameters to optimize. 

To avoid probing regimes which are not physical, we impose that the drive must be off at $t = 0$ and $t = \tau$. We also add a simple smoothing between neighboring steps, as discussed in Sec.~\ref{sec:drive_pulse}. Finally, we impose a strict upper bound on the amplitude of each step as $|\Omega_d^{\mathrm{max}}|/2\pi < 400$ MHz. Due to a different definition of $\Omega_d(t)$, this upper bound coincides with the one used in Ref.~\cite{gautier_optimal_2025}, as detailed above.

The drive pulse is initialized in the square pulse shape, i.e. with all real pulse steps equal and of amplitude $\Omega_d / 2\pi = 150$ MHz while all imaginary steps are zero. Each optimization step is then performed using a Simultaneous Perturbation Stochastic Approximation algorithm which approximates the full gradient as a finite difference directional derivative in a random direction~\cite{spall_multivariate_1992, Bhatnagar_Prasad_Prashanth_2013}. Each step then requires only two function calls, which in turn each require two evolutions: one each for $\beta_g$ and $\beta_e$. These evolutions are performed with the LRA for a rank of $M = 20$. The optimized results are validated with a single additional simulation in an enlarged Hilbert space, with the full Lindblad equation Eq.~\eqref{eq:hamiltonian}. For these simulations, we also increase the rank to $M = 80$ in order to further verify the convergence of our final results with respect to the rank. We find no significant difference in the final pulse shapes when performing the optimization with $M = 80$ as opposed to the faster $M=20$. The results of this optimization with $M = 80$ are displayed in Fig.~\ref{fig:optimization_80}. 

Finally, Fig.~\ref{fig:SNR_Ms} reports the dependence of the assignment error on $M$. The relative error between the LRA and exact values decreases from 5\% to 1.5\% when increasing $M$ from 20 to 80, confirming the accuracy of the controlled LRA.

%% file: lrreadout.bbl
\providecommand{\noopsort}[1]{}\providecommand{\singleletter}[1]{#1}%
\begin{thebibliography}{86}
\expandafter\ifx\csname natexlab\endcsname\relax\def\natexlab#1{#1}\fi
\expandafter\ifx\csname bibnamefont\endcsname\relax
  \def\bibnamefont#1{#1}\fi
\expandafter\ifx\csname bibfnamefont\endcsname\relax
  \def\bibfnamefont#1{#1}\fi
\expandafter\ifx\csname citenamefont\endcsname\relax
  \def\citenamefont#1{#1}\fi
\expandafter\ifx\csname url\endcsname\relax
  \def\url#1{\texttt{#1}}\fi
\expandafter\ifx\csname urlprefix\endcsname\relax\def\urlprefix{URL }\fi
\providecommand{\bibinfo}[2]{#2}
\providecommand{\eprint}[2][]{\url{#2}}

\bibitem[{\citenamefont{Peirce et~al.}(1988)\citenamefont{Peirce, Dahleh, and Rabitz}}]{peirce_optimal_1988}
\bibinfo{author}{\bibfnamefont{A.~P.} \bibnamefont{Peirce}}, \bibinfo{author}{\bibfnamefont{M.~A.} \bibnamefont{Dahleh}}, \bibnamefont{and} \bibinfo{author}{\bibfnamefont{H.}~\bibnamefont{Rabitz}}, \bibinfo{journal}{Physical Review A} \textbf{\bibinfo{volume}{37}}, \bibinfo{pages}{4950} (\bibinfo{year}{1988}), \bibinfo{note}{publisher: American Physical Society}, \urlprefix\url{https://link.aps.org/doi/10.1103/PhysRevA.37.4950}.

\bibitem[{\citenamefont{Werschnik and Gross}(2007)}]{werschnik_quantum_2007}
\bibinfo{author}{\bibfnamefont{J.}~\bibnamefont{Werschnik}} \bibnamefont{and} \bibinfo{author}{\bibfnamefont{E.~K.~U.} \bibnamefont{Gross}}, \bibinfo{journal}{Journal of Physics B: Atomic, Molecular and Optical Physics} \textbf{\bibinfo{volume}{40}}, \bibinfo{pages}{R175} (\bibinfo{year}{2007}), ISSN \bibinfo{issn}{0953-4075, 1361-6455}, \urlprefix\url{https://iopscience.iop.org/article/10.1088/0953-4075/40/18/R01}.

\bibitem[{\citenamefont{Machnes et~al.}(2011)\citenamefont{Machnes, Sander, Glaser, de~Fouqui\`eres, Gruslys, Schirmer, and Schulte-Herbr\"uggen}}]{Machnes_comparing_2011}
\bibinfo{author}{\bibfnamefont{S.}~\bibnamefont{Machnes}}, \bibinfo{author}{\bibfnamefont{U.}~\bibnamefont{Sander}}, \bibinfo{author}{\bibfnamefont{S.~J.} \bibnamefont{Glaser}}, \bibinfo{author}{\bibfnamefont{P.}~\bibnamefont{de~Fouqui\`eres}}, \bibinfo{author}{\bibfnamefont{A.}~\bibnamefont{Gruslys}}, \bibinfo{author}{\bibfnamefont{S.}~\bibnamefont{Schirmer}}, \bibnamefont{and} \bibinfo{author}{\bibfnamefont{T.}~\bibnamefont{Schulte-Herbr\"uggen}}, \bibinfo{journal}{Phys. Rev. A} \textbf{\bibinfo{volume}{84}}, \bibinfo{pages}{022305} (\bibinfo{year}{2011}), \urlprefix\url{https://link.aps.org/doi/10.1103/PhysRevA.84.022305}.

\bibitem[{\citenamefont{Koch et~al.}(2022)\citenamefont{Koch, Boscain, Calarco, Dirr, Filipp, Glaser, Kosloff, Montangero, Schulte-Herbrüggen, Sugny et~al.}}]{koch_quantum_2022}
\bibinfo{author}{\bibfnamefont{C.~P.} \bibnamefont{Koch}}, \bibinfo{author}{\bibfnamefont{U.}~\bibnamefont{Boscain}}, \bibinfo{author}{\bibfnamefont{T.}~\bibnamefont{Calarco}}, \bibinfo{author}{\bibfnamefont{G.}~\bibnamefont{Dirr}}, \bibinfo{author}{\bibfnamefont{S.}~\bibnamefont{Filipp}}, \bibinfo{author}{\bibfnamefont{S.~J.} \bibnamefont{Glaser}}, \bibinfo{author}{\bibfnamefont{R.}~\bibnamefont{Kosloff}}, \bibinfo{author}{\bibfnamefont{S.}~\bibnamefont{Montangero}}, \bibinfo{author}{\bibfnamefont{T.}~\bibnamefont{Schulte-Herbrüggen}}, \bibinfo{author}{\bibfnamefont{D.}~\bibnamefont{Sugny}}, \bibnamefont{et~al.}, \bibinfo{journal}{EPJ Quantum Technology} \textbf{\bibinfo{volume}{9}}, \bibinfo{pages}{1} (\bibinfo{year}{2022}), ISSN \bibinfo{issn}{2196-0763}, \bibinfo{note}{publisher: SpringerOpen}, \urlprefix\url{https://epjquantumtechnology.springeropen.com/articles/10.1140/epjqt/s40507-022-00138-x}.

\bibitem[{\citenamefont{Koch}(2016)}]{koch_controlling_2016}
\bibinfo{author}{\bibfnamefont{C.~P.} \bibnamefont{Koch}}, \bibinfo{journal}{Journal of Physics: Condensed Matter} \textbf{\bibinfo{volume}{28}}, \bibinfo{pages}{213001} (\bibinfo{year}{2016}), ISSN \bibinfo{issn}{0953-8984, 1361-648X}, \urlprefix\url{https://iopscience.iop.org/article/10.1088/0953-8984/28/21/213001}.

\bibitem[{\citenamefont{Schmidt et~al.}(2011)\citenamefont{Schmidt, Negretti, Ankerhold, Calarco, and Stockburger}}]{schmidt_optimal_2011}
\bibinfo{author}{\bibfnamefont{R.}~\bibnamefont{Schmidt}}, \bibinfo{author}{\bibfnamefont{A.}~\bibnamefont{Negretti}}, \bibinfo{author}{\bibfnamefont{J.}~\bibnamefont{Ankerhold}}, \bibinfo{author}{\bibfnamefont{T.}~\bibnamefont{Calarco}}, \bibnamefont{and} \bibinfo{author}{\bibfnamefont{J.~T.} \bibnamefont{Stockburger}}, \bibinfo{journal}{Physical Review Letters} \textbf{\bibinfo{volume}{107}}, \bibinfo{pages}{130404} (\bibinfo{year}{2011}), \bibinfo{note}{publisher: American Physical Society}, \urlprefix\url{https://link.aps.org/doi/10.1103/PhysRevLett.107.130404}.

\bibitem[{\citenamefont{Sauvage and Mintert}(2022)}]{sauvage_optimal_2022}
\bibinfo{author}{\bibfnamefont{F.}~\bibnamefont{Sauvage}} \bibnamefont{and} \bibinfo{author}{\bibfnamefont{F.}~\bibnamefont{Mintert}}, \bibinfo{journal}{Physical Review Letters} \textbf{\bibinfo{volume}{129}}, \bibinfo{pages}{050507} (\bibinfo{year}{2022}), \bibinfo{note}{publisher: American Physical Society}, \urlprefix\url{https://link.aps.org/doi/10.1103/PhysRevLett.129.050507}.

\bibitem[{\citenamefont{Sarma and Hartmann}(2025)}]{sarma_designing_2025}
\bibinfo{author}{\bibfnamefont{B.}~\bibnamefont{Sarma}} \bibnamefont{and} \bibinfo{author}{\bibfnamefont{M.~J.} \bibnamefont{Hartmann}}, \bibinfo{journal}{Physical Review Applied} \textbf{\bibinfo{volume}{23}}, \bibinfo{pages}{014015} (\bibinfo{year}{2025}), ISSN \bibinfo{issn}{2331-7019}, \bibinfo{note}{arXiv:2312.16358 [quant-ph]}, \urlprefix\url{http://arxiv.org/abs/2312.16358}.

\bibitem[{\citenamefont{Jandura and Pupillo}(2022)}]{jandura_time-optimal_2022}
\bibinfo{author}{\bibfnamefont{S.}~\bibnamefont{Jandura}} \bibnamefont{and} \bibinfo{author}{\bibfnamefont{G.}~\bibnamefont{Pupillo}}, \bibinfo{journal}{Quantum} \textbf{\bibinfo{volume}{6}}, \bibinfo{pages}{712} (\bibinfo{year}{2022}), \bibinfo{note}{publisher: Verein zur Förderung des Open Access Publizierens in den Quantenwissenschaften}, \urlprefix\url{https://quantum-journal.org/papers/q-2022-05-13-712/}.

\bibitem[{\citenamefont{Schulte-Herbrueggen et~al.}(2011)\citenamefont{Schulte-Herbrueggen, Spoerl, Khaneja, and Glaser}}]{schulte-herbrueggen_optimal_2011}
\bibinfo{author}{\bibfnamefont{T.}~\bibnamefont{Schulte-Herbrueggen}}, \bibinfo{author}{\bibfnamefont{A.}~\bibnamefont{Spoerl}}, \bibinfo{author}{\bibfnamefont{N.}~\bibnamefont{Khaneja}}, \bibnamefont{and} \bibinfo{author}{\bibfnamefont{S.~J.} \bibnamefont{Glaser}}, \bibinfo{journal}{Journal of Physics B: Atomic, Molecular and Optical Physics} \textbf{\bibinfo{volume}{44}}, \bibinfo{pages}{154013} (\bibinfo{year}{2011}), ISSN \bibinfo{issn}{0953-4075, 1361-6455}, \bibinfo{note}{arXiv:quant-ph/0609037}, \urlprefix\url{http://arxiv.org/abs/quant-ph/0609037}.

\bibitem[{\citenamefont{Werninghaus et~al.}(2021)\citenamefont{Werninghaus, Egger, Roy, Machnes, Wilhelm, and Filipp}}]{werninghaus_leakage_2021}
\bibinfo{author}{\bibfnamefont{M.}~\bibnamefont{Werninghaus}}, \bibinfo{author}{\bibfnamefont{D.~J.} \bibnamefont{Egger}}, \bibinfo{author}{\bibfnamefont{F.}~\bibnamefont{Roy}}, \bibinfo{author}{\bibfnamefont{S.}~\bibnamefont{Machnes}}, \bibinfo{author}{\bibfnamefont{F.~K.} \bibnamefont{Wilhelm}}, \bibnamefont{and} \bibinfo{author}{\bibfnamefont{S.}~\bibnamefont{Filipp}}, \bibinfo{journal}{npj Quantum Information} \textbf{\bibinfo{volume}{7}}, \bibinfo{pages}{14} (\bibinfo{year}{2021}), ISSN \bibinfo{issn}{2056-6387}, \bibinfo{note}{publisher: Nature Publishing Group}, \urlprefix\url{https://www.nature.com/articles/s41534-020-00346-2}.

\bibitem[{\citenamefont{Gautier et~al.}(2023)\citenamefont{Gautier, Mirrahimi, and Sarlette}}]{gautier_designing_2023}
\bibinfo{author}{\bibfnamefont{R.}~\bibnamefont{Gautier}}, \bibinfo{author}{\bibfnamefont{M.}~\bibnamefont{Mirrahimi}}, \bibnamefont{and} \bibinfo{author}{\bibfnamefont{A.}~\bibnamefont{Sarlette}}, \bibinfo{journal}{PRX Quantum} \textbf{\bibinfo{volume}{4}}, \bibinfo{pages}{040316} (\bibinfo{year}{2023}), \bibinfo{note}{publisher: American Physical Society}, \urlprefix\url{https://link.aps.org/doi/10.1103/PRXQuantum.4.040316}.

\bibitem[{\citenamefont{Propson et~al.}(2022)\citenamefont{Propson, Jackson, Koch, Manchester, and Schuster}}]{propson_robust_2022}
\bibinfo{author}{\bibfnamefont{T.}~\bibnamefont{Propson}}, \bibinfo{author}{\bibfnamefont{B.~E.} \bibnamefont{Jackson}}, \bibinfo{author}{\bibfnamefont{J.}~\bibnamefont{Koch}}, \bibinfo{author}{\bibfnamefont{Z.}~\bibnamefont{Manchester}}, \bibnamefont{and} \bibinfo{author}{\bibfnamefont{D.~I.} \bibnamefont{Schuster}}, \bibinfo{journal}{Physical Review Applied} \textbf{\bibinfo{volume}{17}}, \bibinfo{pages}{014036} (\bibinfo{year}{2022}), \bibinfo{note}{publisher: American Physical Society}, \urlprefix\url{https://link.aps.org/doi/10.1103/PhysRevApplied.17.014036}.

\bibitem[{\citenamefont{Blümel et~al.}(2021{\natexlab{a}})\citenamefont{Blümel, Grzesiak, Nguyen, Green, Li, Maksymov, Linke, and Nam}}]{blumel_efficient_2021}
\bibinfo{author}{\bibfnamefont{R.}~\bibnamefont{Blümel}}, \bibinfo{author}{\bibfnamefont{N.}~\bibnamefont{Grzesiak}}, \bibinfo{author}{\bibfnamefont{N.~H.} \bibnamefont{Nguyen}}, \bibinfo{author}{\bibfnamefont{A.~M.} \bibnamefont{Green}}, \bibinfo{author}{\bibfnamefont{M.}~\bibnamefont{Li}}, \bibinfo{author}{\bibfnamefont{A.}~\bibnamefont{Maksymov}}, \bibinfo{author}{\bibfnamefont{N.~M.} \bibnamefont{Linke}}, \bibnamefont{and} \bibinfo{author}{\bibfnamefont{Y.}~\bibnamefont{Nam}}, \bibinfo{journal}{Physical Review Letters} \textbf{\bibinfo{volume}{126}}, \bibinfo{pages}{220503} (\bibinfo{year}{2021}{\natexlab{a}}), \bibinfo{note}{publisher: American Physical Society}, \urlprefix\url{https://link.aps.org/doi/10.1103/PhysRevLett.126.220503}.

\bibitem[{\citenamefont{Rahman et~al.}(2024)\citenamefont{Rahman, Egger, and Arenz}}]{rahman_learning_2024}
\bibinfo{author}{\bibfnamefont{A.}~\bibnamefont{Rahman}}, \bibinfo{author}{\bibfnamefont{D.~J.} \bibnamefont{Egger}}, \bibnamefont{and} \bibinfo{author}{\bibfnamefont{C.}~\bibnamefont{Arenz}}, \bibinfo{journal}{Physical Review Applied} \textbf{\bibinfo{volume}{22}}, \bibinfo{pages}{054074} (\bibinfo{year}{2024}), \bibinfo{note}{publisher: American Physical Society}, \urlprefix\url{https://link.aps.org/doi/10.1103/PhysRevApplied.22.054074}.

\bibitem[{\citenamefont{Ding et~al.}(2023)\citenamefont{Ding, Hays, Sung, Kannan, An, Di~Paolo, Karamlou, Hazard, Azar, Kim et~al.}}]{ding_high-fidelity_2023}
\bibinfo{author}{\bibfnamefont{L.}~\bibnamefont{Ding}}, \bibinfo{author}{\bibfnamefont{M.}~\bibnamefont{Hays}}, \bibinfo{author}{\bibfnamefont{Y.}~\bibnamefont{Sung}}, \bibinfo{author}{\bibfnamefont{B.}~\bibnamefont{Kannan}}, \bibinfo{author}{\bibfnamefont{J.}~\bibnamefont{An}}, \bibinfo{author}{\bibfnamefont{A.}~\bibnamefont{Di~Paolo}}, \bibinfo{author}{\bibfnamefont{A.~H.} \bibnamefont{Karamlou}}, \bibinfo{author}{\bibfnamefont{T.~M.} \bibnamefont{Hazard}}, \bibinfo{author}{\bibfnamefont{K.}~\bibnamefont{Azar}}, \bibinfo{author}{\bibfnamefont{D.~K.} \bibnamefont{Kim}}, \bibnamefont{et~al.}, \bibinfo{journal}{Physical Review X} \textbf{\bibinfo{volume}{13}}, \bibinfo{pages}{031035} (\bibinfo{year}{2023}), \bibinfo{note}{publisher: American Physical Society}, \urlprefix\url{https://link.aps.org/doi/10.1103/PhysRevX.13.031035}.

\bibitem[{\citenamefont{Hyyppä et~al.}(2024)\citenamefont{Hyyppä, Vepsäläinen, Papič, Chan, Inel, Landra, Liu, Luus, Marxer, Ockeloen-Korppi et~al.}}]{hyyppa_reducing_2024}
\bibinfo{author}{\bibfnamefont{E.}~\bibnamefont{Hyyppä}}, \bibinfo{author}{\bibfnamefont{A.}~\bibnamefont{Vepsäläinen}}, \bibinfo{author}{\bibfnamefont{M.}~\bibnamefont{Papič}}, \bibinfo{author}{\bibfnamefont{C.~F.} \bibnamefont{Chan}}, \bibinfo{author}{\bibfnamefont{S.}~\bibnamefont{Inel}}, \bibinfo{author}{\bibfnamefont{A.}~\bibnamefont{Landra}}, \bibinfo{author}{\bibfnamefont{W.}~\bibnamefont{Liu}}, \bibinfo{author}{\bibfnamefont{J.}~\bibnamefont{Luus}}, \bibinfo{author}{\bibfnamefont{F.}~\bibnamefont{Marxer}}, \bibinfo{author}{\bibfnamefont{C.}~\bibnamefont{Ockeloen-Korppi}}, \bibnamefont{et~al.}, \bibinfo{journal}{PRX Quantum} \textbf{\bibinfo{volume}{5}}, \bibinfo{pages}{030353} (\bibinfo{year}{2024}), \bibinfo{note}{publisher: American Physical Society}, \urlprefix\url{https://link.aps.org/doi/10.1103/PRXQuantum.5.030353}.

\bibitem[{\citenamefont{Sung et~al.}(2021)\citenamefont{Sung, Ding, Braumüller, Vepsäläinen, Kannan, Kjaergaard, Greene, Samach, McNally, Kim et~al.}}]{sung_realization_2021}
\bibinfo{author}{\bibfnamefont{Y.}~\bibnamefont{Sung}}, \bibinfo{author}{\bibfnamefont{L.}~\bibnamefont{Ding}}, \bibinfo{author}{\bibfnamefont{J.}~\bibnamefont{Braumüller}}, \bibinfo{author}{\bibfnamefont{A.}~\bibnamefont{Vepsäläinen}}, \bibinfo{author}{\bibfnamefont{B.}~\bibnamefont{Kannan}}, \bibinfo{author}{\bibfnamefont{M.}~\bibnamefont{Kjaergaard}}, \bibinfo{author}{\bibfnamefont{A.}~\bibnamefont{Greene}}, \bibinfo{author}{\bibfnamefont{G.~O.} \bibnamefont{Samach}}, \bibinfo{author}{\bibfnamefont{C.}~\bibnamefont{McNally}}, \bibinfo{author}{\bibfnamefont{D.}~\bibnamefont{Kim}}, \bibnamefont{et~al.}, \bibinfo{journal}{Physical Review X} \textbf{\bibinfo{volume}{11}}, \bibinfo{pages}{021058} (\bibinfo{year}{2021}), \bibinfo{note}{publisher: American Physical Society}, \urlprefix\url{https://link.aps.org/doi/10.1103/PhysRevX.11.021058}.

\bibitem[{\citenamefont{Tripathi et~al.}(2022)\citenamefont{Tripathi, Chen, Khezri, Yip, Levenson-Falk, and Lidar}}]{tripathi_suppression_2022}
\bibinfo{author}{\bibfnamefont{V.}~\bibnamefont{Tripathi}}, \bibinfo{author}{\bibfnamefont{H.}~\bibnamefont{Chen}}, \bibinfo{author}{\bibfnamefont{M.}~\bibnamefont{Khezri}}, \bibinfo{author}{\bibfnamefont{K.-W.} \bibnamefont{Yip}}, \bibinfo{author}{\bibfnamefont{E.~M.} \bibnamefont{Levenson-Falk}}, \bibnamefont{and} \bibinfo{author}{\bibfnamefont{D.~A.} \bibnamefont{Lidar}}, \bibinfo{journal}{Physical Review Applied} \textbf{\bibinfo{volume}{18}}, \bibinfo{pages}{024068} (\bibinfo{year}{2022}), ISSN \bibinfo{issn}{2331-7019}, \bibinfo{note}{arXiv:2108.04530 [quant-ph]}, \urlprefix\url{http://arxiv.org/abs/2108.04530}.

\bibitem[{\citenamefont{Egger and Wilhelm}(2013)}]{egger_optimized_2013}
\bibinfo{author}{\bibfnamefont{D.~J.} \bibnamefont{Egger}} \bibnamefont{and} \bibinfo{author}{\bibfnamefont{F.~K.} \bibnamefont{Wilhelm}}, \bibinfo{journal}{Superconductor Science and Technology} \textbf{\bibinfo{volume}{27}}, \bibinfo{pages}{014001} (\bibinfo{year}{2013}), ISSN \bibinfo{issn}{0953-2048}, \bibinfo{note}{publisher: IOP Publishing}, \urlprefix\url{https://dx.doi.org/10.1088/0953-2048/27/1/014001}.

\bibitem[{\citenamefont{Blümel et~al.}(2021{\natexlab{b}})\citenamefont{Blümel, Grzesiak, Pisenti, Wright, and Nam}}]{blumel_power-optimal_2021}
\bibinfo{author}{\bibfnamefont{R.}~\bibnamefont{Blümel}}, \bibinfo{author}{\bibfnamefont{N.}~\bibnamefont{Grzesiak}}, \bibinfo{author}{\bibfnamefont{N.}~\bibnamefont{Pisenti}}, \bibinfo{author}{\bibfnamefont{K.}~\bibnamefont{Wright}}, \bibnamefont{and} \bibinfo{author}{\bibfnamefont{Y.}~\bibnamefont{Nam}}, \bibinfo{journal}{npj Quantum Information} \textbf{\bibinfo{volume}{7}}, \bibinfo{pages}{147} (\bibinfo{year}{2021}{\natexlab{b}}), ISSN \bibinfo{issn}{2056-6387}, \bibinfo{note}{publisher: Nature Publishing Group}, \urlprefix\url{https://www.nature.com/articles/s41534-021-00489-w}.

\bibitem[{\citenamefont{Mohan et~al.}(2023)\citenamefont{Mohan, de~Keijzer, and Kokkelmans}}]{mohan_robust_2023}
\bibinfo{author}{\bibfnamefont{M.}~\bibnamefont{Mohan}}, \bibinfo{author}{\bibfnamefont{R.}~\bibnamefont{de~Keijzer}}, \bibnamefont{and} \bibinfo{author}{\bibfnamefont{S.}~\bibnamefont{Kokkelmans}}, \bibinfo{journal}{Physical Review Research} \textbf{\bibinfo{volume}{5}}, \bibinfo{pages}{033052} (\bibinfo{year}{2023}), \bibinfo{note}{publisher: American Physical Society}, \urlprefix\url{https://link.aps.org/doi/10.1103/PhysRevResearch.5.033052}.

\bibitem[{\citenamefont{Kuzmanović et~al.}(2025)\citenamefont{Kuzmanović, Moskalenko, Chang, Stanisavljević, Warren, Hogedal, Aggarwal, Ahmad, Biznárová, Dahiya et~al.}}]{kuzmanovic_neural-network-based_2025}
\bibinfo{author}{\bibfnamefont{M.}~\bibnamefont{Kuzmanović}}, \bibinfo{author}{\bibfnamefont{I.}~\bibnamefont{Moskalenko}}, \bibinfo{author}{\bibfnamefont{Y.-H.} \bibnamefont{Chang}}, \bibinfo{author}{\bibfnamefont{O.}~\bibnamefont{Stanisavljević}}, \bibinfo{author}{\bibfnamefont{C.}~\bibnamefont{Warren}}, \bibinfo{author}{\bibfnamefont{E.}~\bibnamefont{Hogedal}}, \bibinfo{author}{\bibfnamefont{A.}~\bibnamefont{Aggarwal}}, \bibinfo{author}{\bibfnamefont{I.}~\bibnamefont{Ahmad}}, \bibinfo{author}{\bibfnamefont{J.}~\bibnamefont{Biznárová}}, \bibinfo{author}{\bibfnamefont{M.}~\bibnamefont{Dahiya}}, \bibnamefont{et~al.}, \emph{\bibinfo{title}{Neural-network-based design and implementation of fast and robust quantum gates}} (\bibinfo{year}{2025}), \bibinfo{note}{arXiv:2505.02054 [quant-ph]}, \urlprefix\url{http://arxiv.org/abs/2505.02054}.

\bibitem[{\citenamefont{Genois et~al.}(2024)\citenamefont{Genois, Stevenson, Goss, Siddiqi, and Blais}}]{genois_quantum_2024}
\bibinfo{author}{\bibfnamefont{E.}~\bibnamefont{Genois}}, \bibinfo{author}{\bibfnamefont{N.~J.} \bibnamefont{Stevenson}}, \bibinfo{author}{\bibfnamefont{N.}~\bibnamefont{Goss}}, \bibinfo{author}{\bibfnamefont{I.}~\bibnamefont{Siddiqi}}, \bibnamefont{and} \bibinfo{author}{\bibfnamefont{A.}~\bibnamefont{Blais}}, \emph{\bibinfo{title}{Quantum optimal control of superconducting qubits based on machine-learning characterization}} (\bibinfo{year}{2024}), \bibinfo{note}{arXiv:2410.22603 [quant-ph] version: 1}, \urlprefix\url{http://arxiv.org/abs/2410.22603}.

\bibitem[{\citenamefont{Nam~Nguyen et~al.}(2024)\citenamefont{Nam~Nguyen, Motzoi, Metcalf, Birgitta~Whaley, Bukov, and Schmitt}}]{nam_nguyen_reinforcement_2024}
\bibinfo{author}{\bibfnamefont{H.}~\bibnamefont{Nam~Nguyen}}, \bibinfo{author}{\bibfnamefont{F.}~\bibnamefont{Motzoi}}, \bibinfo{author}{\bibfnamefont{M.}~\bibnamefont{Metcalf}}, \bibinfo{author}{\bibfnamefont{K.}~\bibnamefont{Birgitta~Whaley}}, \bibinfo{author}{\bibfnamefont{M.}~\bibnamefont{Bukov}}, \bibnamefont{and} \bibinfo{author}{\bibfnamefont{M.}~\bibnamefont{Schmitt}}, \bibinfo{journal}{Machine Learning: Science and Technology} \textbf{\bibinfo{volume}{5}}, \bibinfo{pages}{025066} (\bibinfo{year}{2024}), ISSN \bibinfo{issn}{2632-2153}, \bibinfo{note}{publisher: IOP Publishing}, \urlprefix\url{https://dx.doi.org/10.1088/2632-2153/ad4f4d}.

\bibitem[{\citenamefont{Heinsoo et~al.}(2018)\citenamefont{Heinsoo, Andersen, Remm, Krinner, Walter, Salathé, Gasparinetti, Besse, Potočnik, Wallraff et~al.}}]{heinsoo_rapid_2018}
\bibinfo{author}{\bibfnamefont{J.}~\bibnamefont{Heinsoo}}, \bibinfo{author}{\bibfnamefont{C.~K.} \bibnamefont{Andersen}}, \bibinfo{author}{\bibfnamefont{A.}~\bibnamefont{Remm}}, \bibinfo{author}{\bibfnamefont{S.}~\bibnamefont{Krinner}}, \bibinfo{author}{\bibfnamefont{T.}~\bibnamefont{Walter}}, \bibinfo{author}{\bibfnamefont{Y.}~\bibnamefont{Salathé}}, \bibinfo{author}{\bibfnamefont{S.}~\bibnamefont{Gasparinetti}}, \bibinfo{author}{\bibfnamefont{J.-C.} \bibnamefont{Besse}}, \bibinfo{author}{\bibfnamefont{A.}~\bibnamefont{Potočnik}}, \bibinfo{author}{\bibfnamefont{A.}~\bibnamefont{Wallraff}}, \bibnamefont{et~al.}, \bibinfo{journal}{Physical Review Applied} \textbf{\bibinfo{volume}{10}}, \bibinfo{pages}{034040} (\bibinfo{year}{2018}), \bibinfo{note}{publisher: American Physical Society}, \urlprefix\url{https://link.aps.org/doi/10.1103/PhysRevApplied.10.034040}.

\bibitem[{\citenamefont{Gautier et~al.}(2025)\citenamefont{Gautier, Genois, and Blais}}]{gautier_optimal_2025}
\bibinfo{author}{\bibfnamefont{R.}~\bibnamefont{Gautier}}, \bibinfo{author}{\bibfnamefont{E.}~\bibnamefont{Genois}}, \bibnamefont{and} \bibinfo{author}{\bibfnamefont{A.}~\bibnamefont{Blais}}, \bibinfo{journal}{Physical Review Letters} \textbf{\bibinfo{volume}{134}}, \bibinfo{pages}{070802} (\bibinfo{year}{2025}), \bibinfo{note}{publisher: American Physical Society}, \urlprefix\url{https://link.aps.org/doi/10.1103/PhysRevLett.134.070802}.

\bibitem[{\citenamefont{Chen et~al.}(2024)\citenamefont{Chen, Fors, Yan, Ali, Abad, Osman, Moschandreou, Lienhard, Kosen, Li et~al.}}]{chen_fast_2024}
\bibinfo{author}{\bibfnamefont{L.}~\bibnamefont{Chen}}, \bibinfo{author}{\bibfnamefont{S.~P.} \bibnamefont{Fors}}, \bibinfo{author}{\bibfnamefont{Z.}~\bibnamefont{Yan}}, \bibinfo{author}{\bibfnamefont{A.}~\bibnamefont{Ali}}, \bibinfo{author}{\bibfnamefont{T.}~\bibnamefont{Abad}}, \bibinfo{author}{\bibfnamefont{A.}~\bibnamefont{Osman}}, \bibinfo{author}{\bibfnamefont{E.}~\bibnamefont{Moschandreou}}, \bibinfo{author}{\bibfnamefont{B.}~\bibnamefont{Lienhard}}, \bibinfo{author}{\bibfnamefont{S.}~\bibnamefont{Kosen}}, \bibinfo{author}{\bibfnamefont{H.-X.} \bibnamefont{Li}}, \bibnamefont{et~al.}, \emph{\bibinfo{title}{Fast unconditional reset and leakage reduction in fixed-frequency transmon qubits}} (\bibinfo{year}{2024}), \bibinfo{note}{arXiv:2409.16748 [quant-ph]}, \urlprefix\url{http://arxiv.org/abs/2409.16748}.

\bibitem[{\citenamefont{Abdelhafez et~al.}(2019)\citenamefont{Abdelhafez, Schuster, and Koch}}]{abdelhafez_gradient-based_2019}
\bibinfo{author}{\bibfnamefont{M.}~\bibnamefont{Abdelhafez}}, \bibinfo{author}{\bibfnamefont{D.~I.} \bibnamefont{Schuster}}, \bibnamefont{and} \bibinfo{author}{\bibfnamefont{J.}~\bibnamefont{Koch}}, \bibinfo{journal}{Physical Review A} \textbf{\bibinfo{volume}{99}}, \bibinfo{pages}{052327} (\bibinfo{year}{2019}), \bibinfo{note}{publisher: American Physical Society}, \urlprefix\url{https://link.aps.org/doi/10.1103/PhysRevA.99.052327}.

\bibitem[{\citenamefont{Gambetta et~al.}(2007)\citenamefont{Gambetta, Braff, Wallraff, Girvin, and Schoelkopf}}]{gambetta_protocols_2007}
\bibinfo{author}{\bibfnamefont{J.}~\bibnamefont{Gambetta}}, \bibinfo{author}{\bibfnamefont{W.~A.} \bibnamefont{Braff}}, \bibinfo{author}{\bibfnamefont{A.}~\bibnamefont{Wallraff}}, \bibinfo{author}{\bibfnamefont{S.~M.} \bibnamefont{Girvin}}, \bibnamefont{and} \bibinfo{author}{\bibfnamefont{R.~J.} \bibnamefont{Schoelkopf}}, \bibinfo{journal}{Physical Review A} \textbf{\bibinfo{volume}{76}}, \bibinfo{pages}{012325} (\bibinfo{year}{2007}), \bibinfo{note}{publisher: American Physical Society}, \urlprefix\url{https://link.aps.org/doi/10.1103/PhysRevA.76.012325}.

\bibitem[{\citenamefont{Kurilovich et~al.}(2025)\citenamefont{Kurilovich, Connolly, Bøttcher, Weiss, Hazra, Joshi, Ding, Nho, Diamond, Kurilovich et~al.}}]{kurilovich_high-frequency_2025}
\bibinfo{author}{\bibfnamefont{P.~D.} \bibnamefont{Kurilovich}}, \bibinfo{author}{\bibfnamefont{T.}~\bibnamefont{Connolly}}, \bibinfo{author}{\bibfnamefont{C.~G.~L.} \bibnamefont{Bøttcher}}, \bibinfo{author}{\bibfnamefont{D.~K.} \bibnamefont{Weiss}}, \bibinfo{author}{\bibfnamefont{S.}~\bibnamefont{Hazra}}, \bibinfo{author}{\bibfnamefont{V.~R.} \bibnamefont{Joshi}}, \bibinfo{author}{\bibfnamefont{A.~Z.} \bibnamefont{Ding}}, \bibinfo{author}{\bibfnamefont{H.}~\bibnamefont{Nho}}, \bibinfo{author}{\bibfnamefont{S.}~\bibnamefont{Diamond}}, \bibinfo{author}{\bibfnamefont{V.~D.} \bibnamefont{Kurilovich}}, \bibnamefont{et~al.}, \emph{\bibinfo{title}{High-frequency readout free from transmon multi-excitation resonances}} (\bibinfo{year}{2025}), \bibinfo{note}{arXiv:2501.09161}, \urlprefix\url{http://arxiv.org/abs/2501.09161}.

\bibitem[{\citenamefont{Chapple et~al.}(2025)\citenamefont{Chapple, Benhayoune-Khadraoui, Richer, and Blais}}]{chapple_balanced_2025}
\bibinfo{author}{\bibfnamefont{A.~A.} \bibnamefont{Chapple}}, \bibinfo{author}{\bibfnamefont{O.}~\bibnamefont{Benhayoune-Khadraoui}}, \bibinfo{author}{\bibfnamefont{S.}~\bibnamefont{Richer}}, \bibnamefont{and} \bibinfo{author}{\bibfnamefont{A.}~\bibnamefont{Blais}}, \emph{\bibinfo{title}{Balanced cross-{Kerr} coupling for superconducting qubit readout}} (\bibinfo{year}{2025}), \bibinfo{note}{arXiv:2501.09010}, \urlprefix\url{http://arxiv.org/abs/2501.09010}.

\bibitem[{\citenamefont{Swiadek et~al.}(2024)\citenamefont{Swiadek, Shillito, Magnard, Remm, Hellings, Lacroix, Ficheux, Zanuz, Norris, Blais et~al.}}]{swiadek_enhancing_2024}
\bibinfo{author}{\bibfnamefont{F.}~\bibnamefont{Swiadek}}, \bibinfo{author}{\bibfnamefont{R.}~\bibnamefont{Shillito}}, \bibinfo{author}{\bibfnamefont{P.}~\bibnamefont{Magnard}}, \bibinfo{author}{\bibfnamefont{A.}~\bibnamefont{Remm}}, \bibinfo{author}{\bibfnamefont{C.}~\bibnamefont{Hellings}}, \bibinfo{author}{\bibfnamefont{N.}~\bibnamefont{Lacroix}}, \bibinfo{author}{\bibfnamefont{Q.}~\bibnamefont{Ficheux}}, \bibinfo{author}{\bibfnamefont{D.~C.} \bibnamefont{Zanuz}}, \bibinfo{author}{\bibfnamefont{G.~J.} \bibnamefont{Norris}}, \bibinfo{author}{\bibfnamefont{A.}~\bibnamefont{Blais}}, \bibnamefont{et~al.}, \bibinfo{journal}{PRX Quantum} \textbf{\bibinfo{volume}{5}}, \bibinfo{pages}{040326} (\bibinfo{year}{2024}), ISSN \bibinfo{issn}{2691-3399}, \urlprefix\url{https://link.aps.org/doi/10.1103/PRXQuantum.5.040326}.

\bibitem[{\citenamefont{Chen et~al.}(2023)\citenamefont{Chen, Li, Lu, Warren, Križan, Kosen, Rommel, Ahmed, Osman, Biznárová et~al.}}]{chen_transmon_2023}
\bibinfo{author}{\bibfnamefont{L.}~\bibnamefont{Chen}}, \bibinfo{author}{\bibfnamefont{H.-X.} \bibnamefont{Li}}, \bibinfo{author}{\bibfnamefont{Y.}~\bibnamefont{Lu}}, \bibinfo{author}{\bibfnamefont{C.~W.} \bibnamefont{Warren}}, \bibinfo{author}{\bibfnamefont{C.~J.} \bibnamefont{Križan}}, \bibinfo{author}{\bibfnamefont{S.}~\bibnamefont{Kosen}}, \bibinfo{author}{\bibfnamefont{M.}~\bibnamefont{Rommel}}, \bibinfo{author}{\bibfnamefont{S.}~\bibnamefont{Ahmed}}, \bibinfo{author}{\bibfnamefont{A.}~\bibnamefont{Osman}}, \bibinfo{author}{\bibfnamefont{J.}~\bibnamefont{Biznárová}}, \bibnamefont{et~al.}, \bibinfo{journal}{npj Quantum Information} \textbf{\bibinfo{volume}{9}}, \bibinfo{pages}{26} (\bibinfo{year}{2023}), ISSN \bibinfo{issn}{2056-6387}, \bibinfo{note}{arXiv:2208.05879 [quant-ph]}, \urlprefix\url{http://arxiv.org/abs/2208.05879}.

\bibitem[{\citenamefont{Bengtsson et~al.}(2024)\citenamefont{Bengtsson, Opremcak, Khezri, Sank, Bourassa, Satzinger, Hong, Erickson, Lester, Miao et~al.}}]{bengtsson_model-based_2024}
\bibinfo{author}{\bibfnamefont{A.}~\bibnamefont{Bengtsson}}, \bibinfo{author}{\bibfnamefont{A.}~\bibnamefont{Opremcak}}, \bibinfo{author}{\bibfnamefont{M.}~\bibnamefont{Khezri}}, \bibinfo{author}{\bibfnamefont{D.}~\bibnamefont{Sank}}, \bibinfo{author}{\bibfnamefont{A.}~\bibnamefont{Bourassa}}, \bibinfo{author}{\bibfnamefont{K.~J.} \bibnamefont{Satzinger}}, \bibinfo{author}{\bibfnamefont{S.}~\bibnamefont{Hong}}, \bibinfo{author}{\bibfnamefont{C.}~\bibnamefont{Erickson}}, \bibinfo{author}{\bibfnamefont{B.~J.} \bibnamefont{Lester}}, \bibinfo{author}{\bibfnamefont{K.~C.} \bibnamefont{Miao}}, \bibnamefont{et~al.}, \emph{\bibinfo{title}{Model-based {Optimization} of {Superconducting} {Qubit} {Readout}}} (\bibinfo{year}{2024}), \bibinfo{note}{arXiv:2308.02079}, \urlprefix\url{http://arxiv.org/abs/2308.02079}.

\bibitem[{\citenamefont{Chatterjee et~al.}(2025)\citenamefont{Chatterjee, Schwinger, and Gao}}]{chatterjee_enhanced_2025}
\bibinfo{author}{\bibfnamefont{A.}~\bibnamefont{Chatterjee}}, \bibinfo{author}{\bibfnamefont{J.}~\bibnamefont{Schwinger}}, \bibnamefont{and} \bibinfo{author}{\bibfnamefont{Y.~Y.} \bibnamefont{Gao}}, \bibinfo{journal}{Physical Review Applied} \textbf{\bibinfo{volume}{23}}, \bibinfo{pages}{054057} (\bibinfo{year}{2025}), ISSN \bibinfo{issn}{2331-7019}, \urlprefix\url{https://link.aps.org/doi/10.1103/PhysRevApplied.23.054057}.

\bibitem[{\citenamefont{Egger et~al.}(2018)\citenamefont{Egger, Werninghaus, Ganzhorn, Salis, Fuhrer, Müller, and Filipp}}]{egger_pulsed_2018}
\bibinfo{author}{\bibfnamefont{D.}~\bibnamefont{Egger}}, \bibinfo{author}{\bibfnamefont{M.}~\bibnamefont{Werninghaus}}, \bibinfo{author}{\bibfnamefont{M.}~\bibnamefont{Ganzhorn}}, \bibinfo{author}{\bibfnamefont{G.}~\bibnamefont{Salis}}, \bibinfo{author}{\bibfnamefont{A.}~\bibnamefont{Fuhrer}}, \bibinfo{author}{\bibfnamefont{P.}~\bibnamefont{Müller}}, \bibnamefont{and} \bibinfo{author}{\bibfnamefont{S.}~\bibnamefont{Filipp}}, \bibinfo{journal}{Physical Review Applied} \textbf{\bibinfo{volume}{10}}, \bibinfo{pages}{044030} (\bibinfo{year}{2018}), \bibinfo{note}{publisher: American Physical Society}, \urlprefix\url{https://link.aps.org/doi/10.1103/PhysRevApplied.10.044030}.

\bibitem[{\citenamefont{McEwen et~al.}(2021)\citenamefont{McEwen, Kafri, Chen, Atalaya, Satzinger, Quintana, Klimov, Sank, Gidney, Fowler et~al.}}]{mcewen_removing_2021}
\bibinfo{author}{\bibfnamefont{M.}~\bibnamefont{McEwen}}, \bibinfo{author}{\bibfnamefont{D.}~\bibnamefont{Kafri}}, \bibinfo{author}{\bibfnamefont{Z.}~\bibnamefont{Chen}}, \bibinfo{author}{\bibfnamefont{J.}~\bibnamefont{Atalaya}}, \bibinfo{author}{\bibfnamefont{K.~J.} \bibnamefont{Satzinger}}, \bibinfo{author}{\bibfnamefont{C.}~\bibnamefont{Quintana}}, \bibinfo{author}{\bibfnamefont{P.~V.} \bibnamefont{Klimov}}, \bibinfo{author}{\bibfnamefont{D.}~\bibnamefont{Sank}}, \bibinfo{author}{\bibfnamefont{C.}~\bibnamefont{Gidney}}, \bibinfo{author}{\bibfnamefont{A.~G.} \bibnamefont{Fowler}}, \bibnamefont{et~al.}, \bibinfo{journal}{Nature Communications} \textbf{\bibinfo{volume}{12}}, \bibinfo{pages}{1761} (\bibinfo{year}{2021}), ISSN \bibinfo{issn}{2041-1723}, \bibinfo{note}{publisher: Nature Publishing Group}, \urlprefix\url{https://www.nature.com/articles/s41467-021-21982-y}.

\bibitem[{\citenamefont{Klimov et~al.}(2024)\citenamefont{Klimov, Bengtsson, Quintana, Bourassa, Hong, Dunsworth, Satzinger, Livingston, Sivak, Niu et~al.}}]{klimov_optimizing_2024}
\bibinfo{author}{\bibfnamefont{P.~V.} \bibnamefont{Klimov}}, \bibinfo{author}{\bibfnamefont{A.}~\bibnamefont{Bengtsson}}, \bibinfo{author}{\bibfnamefont{C.}~\bibnamefont{Quintana}}, \bibinfo{author}{\bibfnamefont{A.}~\bibnamefont{Bourassa}}, \bibinfo{author}{\bibfnamefont{S.}~\bibnamefont{Hong}}, \bibinfo{author}{\bibfnamefont{A.}~\bibnamefont{Dunsworth}}, \bibinfo{author}{\bibfnamefont{K.~J.} \bibnamefont{Satzinger}}, \bibinfo{author}{\bibfnamefont{W.~P.} \bibnamefont{Livingston}}, \bibinfo{author}{\bibfnamefont{V.}~\bibnamefont{Sivak}}, \bibinfo{author}{\bibfnamefont{M.~Y.} \bibnamefont{Niu}}, \bibnamefont{et~al.}, \bibinfo{journal}{Nature Communications} \textbf{\bibinfo{volume}{15}}, \bibinfo{pages}{2442} (\bibinfo{year}{2024}), ISSN \bibinfo{issn}{2041-1723}, \bibinfo{note}{publisher: Nature Publishing Group}, \urlprefix\url{https://www.nature.com/articles/s41467-024-46623-y}.

\bibitem[{\citenamefont{Klimov et~al.}(2020)\citenamefont{Klimov, Kelly, Martinis, and Neven}}]{klimov_snake_2020}
\bibinfo{author}{\bibfnamefont{P.~V.} \bibnamefont{Klimov}}, \bibinfo{author}{\bibfnamefont{J.}~\bibnamefont{Kelly}}, \bibinfo{author}{\bibfnamefont{J.~M.} \bibnamefont{Martinis}}, \bibnamefont{and} \bibinfo{author}{\bibfnamefont{H.}~\bibnamefont{Neven}}, \emph{\bibinfo{title}{The {Snake} {Optimizer} for {Learning} {Quantum} {Processor} {Control} {Parameters}}} (\bibinfo{year}{2020}), \bibinfo{note}{arXiv:2006.04594 [quant-ph]}, \urlprefix\url{http://arxiv.org/abs/2006.04594}.

\bibitem[{\citenamefont{Sivak et~al.}(2023)\citenamefont{Sivak, Eickbusch, Royer, Singh, Tsioutsios, Ganjam, Miano, Brock, Ding, Frunzio et~al.}}]{sivak_real-time_2023}
\bibinfo{author}{\bibfnamefont{V.~V.} \bibnamefont{Sivak}}, \bibinfo{author}{\bibfnamefont{A.}~\bibnamefont{Eickbusch}}, \bibinfo{author}{\bibfnamefont{B.}~\bibnamefont{Royer}}, \bibinfo{author}{\bibfnamefont{S.}~\bibnamefont{Singh}}, \bibinfo{author}{\bibfnamefont{I.}~\bibnamefont{Tsioutsios}}, \bibinfo{author}{\bibfnamefont{S.}~\bibnamefont{Ganjam}}, \bibinfo{author}{\bibfnamefont{A.}~\bibnamefont{Miano}}, \bibinfo{author}{\bibfnamefont{B.~L.} \bibnamefont{Brock}}, \bibinfo{author}{\bibfnamefont{A.~Z.} \bibnamefont{Ding}}, \bibinfo{author}{\bibfnamefont{L.}~\bibnamefont{Frunzio}}, \bibnamefont{et~al.}, \bibinfo{journal}{Nature} \textbf{\bibinfo{volume}{616}}, \bibinfo{pages}{50} (\bibinfo{year}{2023}), ISSN \bibinfo{issn}{1476-4687}, \bibinfo{note}{publisher: Nature Publishing Group}, \urlprefix\url{https://www.nature.com/articles/s41586-023-05782-6}.

\bibitem[{\citenamefont{Acharya et~al.}(2025)\citenamefont{Acharya, Abanin, Aghababaie-Beni, Aleiner, Andersen, Ansmann, Arute, Arya, Asfaw, Astrakhantsev et~al.}}]{acharya_quantum_2025}
\bibinfo{author}{\bibfnamefont{R.}~\bibnamefont{Acharya}}, \bibinfo{author}{\bibfnamefont{D.~A.} \bibnamefont{Abanin}}, \bibinfo{author}{\bibfnamefont{L.}~\bibnamefont{Aghababaie-Beni}}, \bibinfo{author}{\bibfnamefont{I.}~\bibnamefont{Aleiner}}, \bibinfo{author}{\bibfnamefont{T.~I.} \bibnamefont{Andersen}}, \bibinfo{author}{\bibfnamefont{M.}~\bibnamefont{Ansmann}}, \bibinfo{author}{\bibfnamefont{F.}~\bibnamefont{Arute}}, \bibinfo{author}{\bibfnamefont{K.}~\bibnamefont{Arya}}, \bibinfo{author}{\bibfnamefont{A.}~\bibnamefont{Asfaw}}, \bibinfo{author}{\bibfnamefont{N.}~\bibnamefont{Astrakhantsev}}, \bibnamefont{et~al.}, \bibinfo{journal}{Nature} \textbf{\bibinfo{volume}{638}}, \bibinfo{pages}{920} (\bibinfo{year}{2025}), ISSN \bibinfo{issn}{1476-4687}, \bibinfo{note}{publisher: Nature Publishing Group}, \urlprefix\url{https://www.nature.com/articles/s41586-024-08449-y}.

\bibitem[{\citenamefont{Sivak et~al.}(2024)\citenamefont{Sivak, Newman, and Klimov}}]{sivak_optimization_2024}
\bibinfo{author}{\bibfnamefont{V.}~\bibnamefont{Sivak}}, \bibinfo{author}{\bibfnamefont{M.}~\bibnamefont{Newman}}, \bibnamefont{and} \bibinfo{author}{\bibfnamefont{P.}~\bibnamefont{Klimov}}, \bibinfo{journal}{Physical Review Letters} \textbf{\bibinfo{volume}{133}}, \bibinfo{pages}{150603} (\bibinfo{year}{2024}), \bibinfo{note}{publisher: American Physical Society}, \urlprefix\url{https://link.aps.org/doi/10.1103/PhysRevLett.133.150603}.

\bibitem[{\citenamefont{Acharya et~al.}(2023)\citenamefont{Acharya, Aleiner, Allen, Andersen, Ansmann, Arute, Arya, Asfaw, Atalaya, Babbush et~al.}}]{acharya_suppressing_2023}
\bibinfo{author}{\bibfnamefont{R.}~\bibnamefont{Acharya}}, \bibinfo{author}{\bibfnamefont{I.}~\bibnamefont{Aleiner}}, \bibinfo{author}{\bibfnamefont{R.}~\bibnamefont{Allen}}, \bibinfo{author}{\bibfnamefont{T.~I.} \bibnamefont{Andersen}}, \bibinfo{author}{\bibfnamefont{M.}~\bibnamefont{Ansmann}}, \bibinfo{author}{\bibfnamefont{F.}~\bibnamefont{Arute}}, \bibinfo{author}{\bibfnamefont{K.}~\bibnamefont{Arya}}, \bibinfo{author}{\bibfnamefont{A.}~\bibnamefont{Asfaw}}, \bibinfo{author}{\bibfnamefont{J.}~\bibnamefont{Atalaya}}, \bibinfo{author}{\bibfnamefont{R.}~\bibnamefont{Babbush}}, \bibnamefont{et~al.}, \bibinfo{journal}{Nature} \textbf{\bibinfo{volume}{614}}, \bibinfo{pages}{676} (\bibinfo{year}{2023}), ISSN \bibinfo{issn}{1476-4687}, \bibinfo{note}{publisher: Nature Publishing Group}, \urlprefix\url{https://www.nature.com/articles/s41586-022-05434-1}.

\bibitem[{\citenamefont{Breuer and Petruccione}(2007)}]{breuer_theory_2007}
\bibinfo{author}{\bibfnamefont{H.~P.} \bibnamefont{Breuer}} \bibnamefont{and} \bibinfo{author}{\bibfnamefont{F.}~\bibnamefont{Petruccione}}, \bibinfo{journal}{The Theory of Open Quantum Systems} \textbf{\bibinfo{volume}{9780199213900}}, \bibinfo{pages}{1} (\bibinfo{year}{2007}), \bibinfo{note}{publisher: Oxford University Press ISBN: 9780191706349}.

\bibitem[{\citenamefont{Khaneja et~al.}(2005)\citenamefont{Khaneja, Reiss, Kehlet, Schulte-Herbrüggen, and Glaser}}]{Khaneja_optimal_2005}
\bibinfo{author}{\bibfnamefont{N.}~\bibnamefont{Khaneja}}, \bibinfo{author}{\bibfnamefont{T.}~\bibnamefont{Reiss}}, \bibinfo{author}{\bibfnamefont{C.}~\bibnamefont{Kehlet}}, \bibinfo{author}{\bibfnamefont{T.}~\bibnamefont{Schulte-Herbrüggen}}, \bibnamefont{and} \bibinfo{author}{\bibfnamefont{S.~J.} \bibnamefont{Glaser}}, \bibinfo{journal}{Journal of Magnetic Resonance} \textbf{\bibinfo{volume}{172}}, \bibinfo{pages}{296} (\bibinfo{year}{2005}), ISSN \bibinfo{issn}{1090-7807}, \urlprefix\url{https://www.sciencedirect.com/science/article/pii/S1090780704003696}.

\bibitem[{\citenamefont{Krotov}(1993)}]{Krotov1993}
\bibinfo{author}{\bibfnamefont{V.~F.} \bibnamefont{Krotov}}, \emph{\bibinfo{title}{Global Methods in Optimal Control Theory}} (\bibinfo{publisher}{Birkh{\"a}user Boston}, \bibinfo{address}{Boston, MA}, \bibinfo{year}{1993}), pp. \bibinfo{pages}{74--121}, ISBN \bibinfo{isbn}{978-1-4612-0349-0}, \urlprefix\url{https://doi.org/10.1007/978-1-4612-0349-0_3}.

\bibitem[{\citenamefont{Goerz et~al.}(2019)\citenamefont{Goerz, Basilewitsch, Gago-Encinas, Krauss, Horn, Reich, and Koch}}]{Goerz_Krotov_2019}
\bibinfo{author}{\bibfnamefont{M.~H.} \bibnamefont{Goerz}}, \bibinfo{author}{\bibfnamefont{D.}~\bibnamefont{Basilewitsch}}, \bibinfo{author}{\bibfnamefont{F.}~\bibnamefont{Gago-Encinas}}, \bibinfo{author}{\bibfnamefont{M.~G.} \bibnamefont{Krauss}}, \bibinfo{author}{\bibfnamefont{K.~P.} \bibnamefont{Horn}}, \bibinfo{author}{\bibfnamefont{D.~M.} \bibnamefont{Reich}}, \bibnamefont{and} \bibinfo{author}{\bibfnamefont{C.~P.} \bibnamefont{Koch}}, \bibinfo{journal}{SciPost Phys.} \textbf{\bibinfo{volume}{7}}, \bibinfo{pages}{080} (\bibinfo{year}{2019}), \urlprefix\url{https://scipost.org/10.21468/SciPostPhys.7.6.080}.

\bibitem[{\citenamefont{Caneva et~al.}(2011)\citenamefont{Caneva, Calarco, and Montangero}}]{Caneva_chopped_2011}
\bibinfo{author}{\bibfnamefont{T.}~\bibnamefont{Caneva}}, \bibinfo{author}{\bibfnamefont{T.}~\bibnamefont{Calarco}}, \bibnamefont{and} \bibinfo{author}{\bibfnamefont{S.}~\bibnamefont{Montangero}}, \bibinfo{journal}{Phys. Rev. A} \textbf{\bibinfo{volume}{84}}, \bibinfo{pages}{022326} (\bibinfo{year}{2011}), \urlprefix\url{https://link.aps.org/doi/10.1103/PhysRevA.84.022326}.

\bibitem[{\citenamefont{Müller et~al.}(2022)\citenamefont{Müller, Gherardini, Calarco, Montangero, and Caruso}}]{muller_information_2022}
\bibinfo{author}{\bibfnamefont{M.~M.} \bibnamefont{Müller}}, \bibinfo{author}{\bibfnamefont{S.}~\bibnamefont{Gherardini}}, \bibinfo{author}{\bibfnamefont{T.}~\bibnamefont{Calarco}}, \bibinfo{author}{\bibfnamefont{S.}~\bibnamefont{Montangero}}, \bibnamefont{and} \bibinfo{author}{\bibfnamefont{F.}~\bibnamefont{Caruso}}, \bibinfo{journal}{Scientific Reports} \textbf{\bibinfo{volume}{12}}, \bibinfo{pages}{21405} (\bibinfo{year}{2022}), ISSN \bibinfo{issn}{2045-2322}, \bibinfo{note}{publisher: Nature Publishing Group}, \urlprefix\url{https://www.nature.com/articles/s41598-022-25770-6}.

\bibitem[{\citenamefont{Leung et~al.}(2017)\citenamefont{Leung, Abdelhafez, Koch, and Schuster}}]{leung_speedup_2017}
\bibinfo{author}{\bibfnamefont{N.}~\bibnamefont{Leung}}, \bibinfo{author}{\bibfnamefont{M.}~\bibnamefont{Abdelhafez}}, \bibinfo{author}{\bibfnamefont{J.}~\bibnamefont{Koch}}, \bibnamefont{and} \bibinfo{author}{\bibfnamefont{D.}~\bibnamefont{Schuster}}, \bibinfo{journal}{Physical Review A} \textbf{\bibinfo{volume}{95}}, \bibinfo{pages}{042318} (\bibinfo{year}{2017}), \bibinfo{note}{publisher: American Physical Society}, \urlprefix\url{https://link.aps.org/doi/10.1103/PhysRevA.95.042318}.

\bibitem[{\citenamefont{Boutin et~al.}(2017)\citenamefont{Boutin, Andersen, Venkatraman, Ferris, and Blais}}]{Boutin_resonator_2017}
\bibinfo{author}{\bibfnamefont{S.}~\bibnamefont{Boutin}}, \bibinfo{author}{\bibfnamefont{C.~K.} \bibnamefont{Andersen}}, \bibinfo{author}{\bibfnamefont{J.}~\bibnamefont{Venkatraman}}, \bibinfo{author}{\bibfnamefont{A.~J.} \bibnamefont{Ferris}}, \bibnamefont{and} \bibinfo{author}{\bibfnamefont{A.}~\bibnamefont{Blais}}, \bibinfo{journal}{Phys. Rev. A} \textbf{\bibinfo{volume}{96}}, \bibinfo{pages}{042315} (\bibinfo{year}{2017}), \urlprefix\url{https://link.aps.org/doi/10.1103/PhysRevA.96.042315}.

\bibitem[{\citenamefont{Egger and Wilhelm}(2014)}]{Egger_optimal_2014}
\bibinfo{author}{\bibfnamefont{D.~J.} \bibnamefont{Egger}} \bibnamefont{and} \bibinfo{author}{\bibfnamefont{F.~K.} \bibnamefont{Wilhelm}}, \bibinfo{journal}{Phys. Rev. A} \textbf{\bibinfo{volume}{90}}, \bibinfo{pages}{052331} (\bibinfo{year}{2014}), \urlprefix\url{https://link.aps.org/doi/10.1103/PhysRevA.90.052331}.

\bibitem[{\citenamefont{Le~Bris and Rouchon}(2013)}]{lebris_lowrank_2013}
\bibinfo{author}{\bibfnamefont{C.}~\bibnamefont{Le~Bris}} \bibnamefont{and} \bibinfo{author}{\bibfnamefont{P.}~\bibnamefont{Rouchon}}, \bibinfo{journal}{Phys. Rev. A} \textbf{\bibinfo{volume}{87}}, \bibinfo{pages}{022125} (\bibinfo{year}{2013}), \urlprefix\url{https://link.aps.org/doi/10.1103/PhysRevA.87.022125}.

\bibitem[{\citenamefont{McCaul et~al.}(2021)\citenamefont{McCaul, Jacobs, and Bondar}}]{mccaul_fast_2021}
\bibinfo{author}{\bibfnamefont{G.}~\bibnamefont{McCaul}}, \bibinfo{author}{\bibfnamefont{K.}~\bibnamefont{Jacobs}}, \bibnamefont{and} \bibinfo{author}{\bibfnamefont{D.~I.} \bibnamefont{Bondar}}, \bibinfo{journal}{Phys. Rev. Res.} \textbf{\bibinfo{volume}{3}}, \bibinfo{pages}{013017} (\bibinfo{year}{2021}), \urlprefix\url{https://link.aps.org/doi/10.1103/PhysRevResearch.3.013017}.

\bibitem[{\citenamefont{Chen et~al.}(2021)\citenamefont{Chen, Farquhar, and Parrish}}]{chen_low-rank_2021}
\bibinfo{author}{\bibfnamefont{Y.-T.} \bibnamefont{Chen}}, \bibinfo{author}{\bibfnamefont{C.}~\bibnamefont{Farquhar}}, \bibnamefont{and} \bibinfo{author}{\bibfnamefont{R.~M.} \bibnamefont{Parrish}}, \bibinfo{journal}{npj Quantum Information} \textbf{\bibinfo{volume}{7}}, \bibinfo{pages}{61} (\bibinfo{year}{2021}), ISSN \bibinfo{issn}{2056-6387}, \urlprefix\url{https://doi.org/10.1038/s41534-021-00392-4}.

\bibitem[{\citenamefont{Donatella et~al.}(2021)\citenamefont{Donatella, Denis, Le~Boit\'e, and Ciuti}}]{donatella_continuous-time_2021}
\bibinfo{author}{\bibfnamefont{K.}~\bibnamefont{Donatella}}, \bibinfo{author}{\bibfnamefont{Z.}~\bibnamefont{Denis}}, \bibinfo{author}{\bibfnamefont{A.}~\bibnamefont{Le~Boit\'e}}, \bibnamefont{and} \bibinfo{author}{\bibfnamefont{C.}~\bibnamefont{Ciuti}}, \bibinfo{journal}{Phys. Rev. A} \textbf{\bibinfo{volume}{104}}, \bibinfo{pages}{062407} (\bibinfo{year}{2021}), \urlprefix\url{https://link.aps.org/doi/10.1103/PhysRevA.104.062407}.

\bibitem[{\citenamefont{Santos et~al.}(2025)\citenamefont{Santos, Song, and Savona}}]{santos_low-rank_2025}
\bibinfo{author}{\bibfnamefont{S.}~\bibnamefont{Santos}}, \bibinfo{author}{\bibfnamefont{X.}~\bibnamefont{Song}}, \bibnamefont{and} \bibinfo{author}{\bibfnamefont{V.}~\bibnamefont{Savona}}, \bibinfo{journal}{Quantum} \textbf{\bibinfo{volume}{9}}, \bibinfo{pages}{1620} (\bibinfo{year}{2025}), ISSN \bibinfo{issn}{2521-327X}, \bibinfo{note}{arXiv:2403.05908 [quant-ph]}, \urlprefix\url{http://arxiv.org/abs/2403.05908}.

\bibitem[{\citenamefont{Joubert-Doriol et~al.}(2014)\citenamefont{Joubert-Doriol, Ryabinkin, and Izmaylov}}]{joubert-doriol_non-stochastic_2014}
\bibinfo{author}{\bibfnamefont{L.}~\bibnamefont{Joubert-Doriol}}, \bibinfo{author}{\bibfnamefont{I.~G.} \bibnamefont{Ryabinkin}}, \bibnamefont{and} \bibinfo{author}{\bibfnamefont{A.~F.} \bibnamefont{Izmaylov}}, \bibinfo{journal}{The Journal of Chemical Physics} \textbf{\bibinfo{volume}{141}} (\bibinfo{year}{2014}), ISSN \bibinfo{issn}{0021-9606, 1089-7690}, \urlprefix\url{https://doi.org/10.1063/1.4903829}.

\bibitem[{\citenamefont{Joubert-Doriol and Izmaylov}(2015)}]{joubert-doriol_problem-free_2015}
\bibinfo{author}{\bibfnamefont{L.}~\bibnamefont{Joubert-Doriol}} \bibnamefont{and} \bibinfo{author}{\bibfnamefont{A.~F.} \bibnamefont{Izmaylov}}, \bibinfo{journal}{The Journal of Chemical Physics} \textbf{\bibinfo{volume}{142}}, \bibinfo{pages}{134107} (\bibinfo{year}{2015}), ISSN \bibinfo{issn}{0021-9606, 1089-7690}, \urlprefix\url{https://pubs.aip.org/jcp/article/142/13/134107/901966/Problem-free-time-dependent-variational-principle}.

\bibitem[{\citenamefont{Gravina and Savona}(2024)}]{gravina_adaptive_2024}
\bibinfo{author}{\bibfnamefont{L.}~\bibnamefont{Gravina}} \bibnamefont{and} \bibinfo{author}{\bibfnamefont{V.}~\bibnamefont{Savona}}, \bibinfo{journal}{Physical Review Research} \textbf{\bibinfo{volume}{6}}, \bibinfo{pages}{023072} (\bibinfo{year}{2024}), ISSN \bibinfo{issn}{2643-1564}, \urlprefix\url{https://link.aps.org/doi/10.1103/PhysRevResearch.6.023072}.

\bibitem[{\citenamefont{Blais et~al.}(2004)\citenamefont{Blais, Huang, Wallraff, Girvin, and Schoelkopf}}]{blais_cavity_2004}
\bibinfo{author}{\bibfnamefont{A.}~\bibnamefont{Blais}}, \bibinfo{author}{\bibfnamefont{R.-S.} \bibnamefont{Huang}}, \bibinfo{author}{\bibfnamefont{A.}~\bibnamefont{Wallraff}}, \bibinfo{author}{\bibfnamefont{S.~M.} \bibnamefont{Girvin}}, \bibnamefont{and} \bibinfo{author}{\bibfnamefont{R.~J.} \bibnamefont{Schoelkopf}}, \bibinfo{journal}{Physical Review A} \textbf{\bibinfo{volume}{69}}, \bibinfo{pages}{062320} (\bibinfo{year}{2004}), ISSN \bibinfo{issn}{1050-2947, 1094-1622}, \urlprefix\url{https://link.aps.org/doi/10.1103/PhysRevA.69.062320}.

\bibitem[{\citenamefont{Touzard et~al.}(2019)\citenamefont{Touzard, Kou, Frattini, Sivak, Puri, Grimm, Frunzio, Shankar, and Devoret}}]{touzard_gated_2019}
\bibinfo{author}{\bibfnamefont{S.}~\bibnamefont{Touzard}}, \bibinfo{author}{\bibfnamefont{A.}~\bibnamefont{Kou}}, \bibinfo{author}{\bibfnamefont{N.~E.} \bibnamefont{Frattini}}, \bibinfo{author}{\bibfnamefont{V.~V.} \bibnamefont{Sivak}}, \bibinfo{author}{\bibfnamefont{S.}~\bibnamefont{Puri}}, \bibinfo{author}{\bibfnamefont{A.}~\bibnamefont{Grimm}}, \bibinfo{author}{\bibfnamefont{L.}~\bibnamefont{Frunzio}}, \bibinfo{author}{\bibfnamefont{S.}~\bibnamefont{Shankar}}, \bibnamefont{and} \bibinfo{author}{\bibfnamefont{M.~H.} \bibnamefont{Devoret}}, \bibinfo{journal}{Phys. Rev. Lett.} \textbf{\bibinfo{volume}{122}}, \bibinfo{pages}{080502} (\bibinfo{year}{2019}), \urlprefix\url{https://link.aps.org/doi/10.1103/PhysRevLett.122.080502}.

\bibitem[{\citenamefont{Walter et~al.}(2017)\citenamefont{Walter, Kurpiers, Gasparinetti, Magnard, Potočnik, Salathé, Pechal, Mondal, Oppliger, Eichler et~al.}}]{walter_rapid_2017}
\bibinfo{author}{\bibfnamefont{T.}~\bibnamefont{Walter}}, \bibinfo{author}{\bibfnamefont{P.}~\bibnamefont{Kurpiers}}, \bibinfo{author}{\bibfnamefont{S.}~\bibnamefont{Gasparinetti}}, \bibinfo{author}{\bibfnamefont{P.}~\bibnamefont{Magnard}}, \bibinfo{author}{\bibfnamefont{A.}~\bibnamefont{Potočnik}}, \bibinfo{author}{\bibfnamefont{Y.}~\bibnamefont{Salathé}}, \bibinfo{author}{\bibfnamefont{M.}~\bibnamefont{Pechal}}, \bibinfo{author}{\bibfnamefont{M.}~\bibnamefont{Mondal}}, \bibinfo{author}{\bibfnamefont{M.}~\bibnamefont{Oppliger}}, \bibinfo{author}{\bibfnamefont{C.}~\bibnamefont{Eichler}}, \bibnamefont{et~al.}, \bibinfo{journal}{Physical Review Applied} \textbf{\bibinfo{volume}{7}}, \bibinfo{pages}{054020} (\bibinfo{year}{2017}), ISSN \bibinfo{issn}{2331-7019}, \urlprefix\url{http://link.aps.org/doi/10.1103/PhysRevApplied.7.054020}.

\bibitem[{\citenamefont{Sunada et~al.}(2022)\citenamefont{Sunada, Kono, Ilves, Tamate, Sugiyama, Tabuchi, and Nakamura}}]{sunada_fast_2022}
\bibinfo{author}{\bibfnamefont{Y.}~\bibnamefont{Sunada}}, \bibinfo{author}{\bibfnamefont{S.}~\bibnamefont{Kono}}, \bibinfo{author}{\bibfnamefont{J.}~\bibnamefont{Ilves}}, \bibinfo{author}{\bibfnamefont{S.}~\bibnamefont{Tamate}}, \bibinfo{author}{\bibfnamefont{T.}~\bibnamefont{Sugiyama}}, \bibinfo{author}{\bibfnamefont{Y.}~\bibnamefont{Tabuchi}}, \bibnamefont{and} \bibinfo{author}{\bibfnamefont{Y.}~\bibnamefont{Nakamura}}, \bibinfo{journal}{Physical Review Applied} \textbf{\bibinfo{volume}{17}}, \bibinfo{pages}{044016} (\bibinfo{year}{2022}), ISSN \bibinfo{issn}{2331-7019}, \urlprefix\url{https://link.aps.org/doi/10.1103/PhysRevApplied.17.044016}.

\bibitem[{\citenamefont{Reed et~al.}(2010)\citenamefont{Reed, Johnson, Houck, DiCarlo, Chow, Schuster, Frunzio, and Schoelkopf}}]{reed_fast_2010}
\bibinfo{author}{\bibfnamefont{M.~D.} \bibnamefont{Reed}}, \bibinfo{author}{\bibfnamefont{B.~R.} \bibnamefont{Johnson}}, \bibinfo{author}{\bibfnamefont{A.~A.} \bibnamefont{Houck}}, \bibinfo{author}{\bibfnamefont{L.}~\bibnamefont{DiCarlo}}, \bibinfo{author}{\bibfnamefont{J.~M.} \bibnamefont{Chow}}, \bibinfo{author}{\bibfnamefont{D.~I.} \bibnamefont{Schuster}}, \bibinfo{author}{\bibfnamefont{L.}~\bibnamefont{Frunzio}}, \bibnamefont{and} \bibinfo{author}{\bibfnamefont{R.~J.} \bibnamefont{Schoelkopf}}, \bibinfo{journal}{Applied Physics Letters} \textbf{\bibinfo{volume}{96}}, \bibinfo{pages}{203110} (\bibinfo{year}{2010}), ISSN \bibinfo{issn}{0003-6951}, \urlprefix\url{https://doi.org/10.1063/1.3435463}.

\bibitem[{\citenamefont{Blais et~al.}(2021)\citenamefont{Blais, Grimsmo, Girvin, and Wallraff}}]{blais_circuit_2021}
\bibinfo{author}{\bibfnamefont{A.}~\bibnamefont{Blais}}, \bibinfo{author}{\bibfnamefont{A.~L.} \bibnamefont{Grimsmo}}, \bibinfo{author}{\bibfnamefont{S.}~\bibnamefont{Girvin}}, \bibnamefont{and} \bibinfo{author}{\bibfnamefont{A.}~\bibnamefont{Wallraff}}, \bibinfo{journal}{Reviews of Modern Physics} \textbf{\bibinfo{volume}{93}}, \bibinfo{pages}{025005} (\bibinfo{year}{2021}), ISSN \bibinfo{issn}{0034-6861, 1539-0756}, \urlprefix\url{https://link.aps.org/doi/10.1103/RevModPhys.93.025005}.

\bibitem[{SM()}]{SM}
\bibinfo{note}{See Supplemental Material for more information.}

\bibitem[{\citenamefont{Goutte}(2025)}]{goutte_github_2025}
\bibinfo{author}{\bibfnamefont{L.}~\bibnamefont{Goutte}}, \emph{\bibinfo{title}{github.com/leogoutte/low\_rank}} (\bibinfo{year}{2025}), \urlprefix\url{https://github.com/leogoutte/low\_rank}.

\bibitem[{\citenamefont{Mercurio et~al.}(2025)\citenamefont{Mercurio, Huang, Cai, Chen, Savona, and Nori}}]{mercurio_quantumtoolboxjl_2025}
\bibinfo{author}{\bibfnamefont{A.}~\bibnamefont{Mercurio}}, \bibinfo{author}{\bibfnamefont{Y.-T.} \bibnamefont{Huang}}, \bibinfo{author}{\bibfnamefont{L.-X.} \bibnamefont{Cai}}, \bibinfo{author}{\bibfnamefont{Y.-N.} \bibnamefont{Chen}}, \bibinfo{author}{\bibfnamefont{V.}~\bibnamefont{Savona}}, \bibnamefont{and} \bibinfo{author}{\bibfnamefont{F.}~\bibnamefont{Nori}}, \bibinfo{journal}{{Quantum}} \textbf{\bibinfo{volume}{9}}, \bibinfo{pages}{1866} (\bibinfo{year}{2025}), ISSN \bibinfo{issn}{2521-327X}, \urlprefix\url{https://doi.org/10.22331/q-2025-09-29-1866}.

\bibitem[{\citenamefont{Spall}(1992)}]{spall_multivariate_1992}
\bibinfo{author}{\bibfnamefont{J.}~\bibnamefont{Spall}}, \bibinfo{journal}{IEEE Transactions on Automatic Control} \textbf{\bibinfo{volume}{37}}, \bibinfo{pages}{332} (\bibinfo{year}{1992}), ISSN \bibinfo{issn}{1558-2523}, \urlprefix\url{https://ieeexplore.ieee.org/document/119632/}.

\bibitem[{\citenamefont{Bhatnagar et~al.}(2013)\citenamefont{Bhatnagar, Prasad, and Prashanth}}]{Bhatnagar_Prasad_Prashanth_2013}
\bibinfo{author}{\bibfnamefont{S.}~\bibnamefont{Bhatnagar}}, \bibinfo{author}{\bibfnamefont{H.~L.} \bibnamefont{Prasad}}, \bibnamefont{and} \bibinfo{author}{\bibfnamefont{L.~A.} \bibnamefont{Prashanth}}, \emph{\bibinfo{title}{Stochastic recursive algorithms for optimization: Simultaneous perturbation methods}} (\bibinfo{publisher}{Springer}, \bibinfo{year}{2013}).

\bibitem[{\citenamefont{Kennedy and Eberhart}(1995)}]{kennedy_pso_1995}
\bibinfo{author}{\bibfnamefont{J.}~\bibnamefont{Kennedy}} \bibnamefont{and} \bibinfo{author}{\bibfnamefont{R.}~\bibnamefont{Eberhart}}, in \emph{\bibinfo{booktitle}{Proceedings of ICNN'95 - International Conference on Neural Networks}} (\bibinfo{year}{1995}), vol.~\bibinfo{volume}{4}, pp. \bibinfo{pages}{1942--1948 vol.4}.

\bibitem[{\citenamefont{McClure et~al.}(2016)\citenamefont{McClure, Paik, Bishop, Steffen, Chow, and Gambetta}}]{mcclure_rapid_2016}
\bibinfo{author}{\bibfnamefont{D.~T.} \bibnamefont{McClure}}, \bibinfo{author}{\bibfnamefont{H.}~\bibnamefont{Paik}}, \bibinfo{author}{\bibfnamefont{L.~S.} \bibnamefont{Bishop}}, \bibinfo{author}{\bibfnamefont{M.}~\bibnamefont{Steffen}}, \bibinfo{author}{\bibfnamefont{J.~M.} \bibnamefont{Chow}}, \bibnamefont{and} \bibinfo{author}{\bibfnamefont{J.~M.} \bibnamefont{Gambetta}}, \bibinfo{journal}{Phys. Rev. Appl.} \textbf{\bibinfo{volume}{5}}, \bibinfo{pages}{011001} (\bibinfo{year}{2016}), \urlprefix\url{https://link.aps.org/doi/10.1103/PhysRevApplied.5.011001}.

\bibitem[{\citenamefont{Mu\~noz Arias et~al.}(2023)\citenamefont{Mu\~noz Arias, Lled\'o, and Blais}}]{arias_qubit_2023}
\bibinfo{author}{\bibfnamefont{M.~H.} \bibnamefont{Mu\~noz Arias}}, \bibinfo{author}{\bibfnamefont{C.}~\bibnamefont{Lled\'o}}, \bibnamefont{and} \bibinfo{author}{\bibfnamefont{A.}~\bibnamefont{Blais}}, \bibinfo{journal}{Phys. Rev. Appl.} \textbf{\bibinfo{volume}{20}}, \bibinfo{pages}{054013} (\bibinfo{year}{2023}), \urlprefix\url{https://link.aps.org/doi/10.1103/PhysRevApplied.20.054013}.

\bibitem[{\citenamefont{Ikonen et~al.}(2019)\citenamefont{Ikonen, Goetz, Ilves, Ker\"anen, Gunyho, Partanen, Tan, Hazra, Gr\"onberg, Vesterinen et~al.}}]{ikonen_qubit_2019}
\bibinfo{author}{\bibfnamefont{J.}~\bibnamefont{Ikonen}}, \bibinfo{author}{\bibfnamefont{J.}~\bibnamefont{Goetz}}, \bibinfo{author}{\bibfnamefont{J.}~\bibnamefont{Ilves}}, \bibinfo{author}{\bibfnamefont{A.}~\bibnamefont{Ker\"anen}}, \bibinfo{author}{\bibfnamefont{A.~M.} \bibnamefont{Gunyho}}, \bibinfo{author}{\bibfnamefont{M.}~\bibnamefont{Partanen}}, \bibinfo{author}{\bibfnamefont{K.~Y.} \bibnamefont{Tan}}, \bibinfo{author}{\bibfnamefont{D.}~\bibnamefont{Hazra}}, \bibinfo{author}{\bibfnamefont{L.}~\bibnamefont{Gr\"onberg}}, \bibinfo{author}{\bibfnamefont{V.}~\bibnamefont{Vesterinen}}, \bibnamefont{et~al.}, \bibinfo{journal}{Phys. Rev. Lett.} \textbf{\bibinfo{volume}{122}}, \bibinfo{pages}{080503} (\bibinfo{year}{2019}), \urlprefix\url{https://link.aps.org/doi/10.1103/PhysRevLett.122.080503}.

\bibitem[{\citenamefont{Manzano}(2020)}]{manzano_lindblad_2020}
\bibinfo{author}{\bibfnamefont{D.}~\bibnamefont{Manzano}}, \bibinfo{journal}{AIP Advances} \textbf{\bibinfo{volume}{10}}, \bibinfo{pages}{025106} (\bibinfo{year}{2020}).

\bibitem[{\citenamefont{Khezri et~al.}(2023)\citenamefont{Khezri, Opremcak, Chen, Miao, McEwen, Bengtsson, White, Naaman, Sank, Korotkov et~al.}}]{khezri_measurement-induced_2023}
\bibinfo{author}{\bibfnamefont{M.}~\bibnamefont{Khezri}}, \bibinfo{author}{\bibfnamefont{A.}~\bibnamefont{Opremcak}}, \bibinfo{author}{\bibfnamefont{Z.}~\bibnamefont{Chen}}, \bibinfo{author}{\bibfnamefont{K.~C.} \bibnamefont{Miao}}, \bibinfo{author}{\bibfnamefont{M.}~\bibnamefont{McEwen}}, \bibinfo{author}{\bibfnamefont{A.}~\bibnamefont{Bengtsson}}, \bibinfo{author}{\bibfnamefont{T.}~\bibnamefont{White}}, \bibinfo{author}{\bibfnamefont{O.}~\bibnamefont{Naaman}}, \bibinfo{author}{\bibfnamefont{D.}~\bibnamefont{Sank}}, \bibinfo{author}{\bibfnamefont{A.~N.} \bibnamefont{Korotkov}}, \bibnamefont{et~al.}, \bibinfo{journal}{Physical Review Applied} \textbf{\bibinfo{volume}{20}}, \bibinfo{pages}{054008} (\bibinfo{year}{2023}), ISSN \bibinfo{issn}{2331-7019}, \urlprefix\url{https://link.aps.org/doi/10.1103/PhysRevApplied.20.054008}.

\bibitem[{\citenamefont{Kofman and Kurizki}(2004)}]{kofman_unified_2004}
\bibinfo{author}{\bibfnamefont{A.~G.} \bibnamefont{Kofman}} \bibnamefont{and} \bibinfo{author}{\bibfnamefont{G.}~\bibnamefont{Kurizki}}, \bibinfo{journal}{Physical Review Letters} \textbf{\bibinfo{volume}{93}}, \bibinfo{pages}{130406} (\bibinfo{year}{2004}), \bibinfo{note}{publisher: American Physical Society}, \urlprefix\url{https://link.aps.org/doi/10.1103/PhysRevLett.93.130406}.

\bibitem[{\citenamefont{Shillito et~al.}(2022)\citenamefont{Shillito, Petrescu, Cohen, Beall, Hauru, Ganahl, Lewis, Vidal, and Blais}}]{shillito_dynamics_2022}
\bibinfo{author}{\bibfnamefont{R.}~\bibnamefont{Shillito}}, \bibinfo{author}{\bibfnamefont{A.}~\bibnamefont{Petrescu}}, \bibinfo{author}{\bibfnamefont{J.}~\bibnamefont{Cohen}}, \bibinfo{author}{\bibfnamefont{J.}~\bibnamefont{Beall}}, \bibinfo{author}{\bibfnamefont{M.}~\bibnamefont{Hauru}}, \bibinfo{author}{\bibfnamefont{M.}~\bibnamefont{Ganahl}}, \bibinfo{author}{\bibfnamefont{A.~G.~M.} \bibnamefont{Lewis}}, \bibinfo{author}{\bibfnamefont{G.}~\bibnamefont{Vidal}}, \bibnamefont{and} \bibinfo{author}{\bibfnamefont{A.}~\bibnamefont{Blais}}, \emph{\bibinfo{title}{Dynamics of {Transmon} {Ionization}}} (\bibinfo{year}{2022}), \bibinfo{note}{arXiv:2203.11235}, \urlprefix\url{http://arxiv.org/abs/2203.11235}.

\bibitem[{\citenamefont{Dumas et~al.}(2024)\citenamefont{Dumas, Groleau-Paré, McDonald, Muñoz-Arias, Lledó, D’Anjou, and Blais}}]{dumas_measurement-induced_2024}
\bibinfo{author}{\bibfnamefont{M.~F.} \bibnamefont{Dumas}}, \bibinfo{author}{\bibfnamefont{B.}~\bibnamefont{Groleau-Paré}}, \bibinfo{author}{\bibfnamefont{A.}~\bibnamefont{McDonald}}, \bibinfo{author}{\bibfnamefont{M.~H.} \bibnamefont{Muñoz-Arias}}, \bibinfo{author}{\bibfnamefont{C.}~\bibnamefont{Lledó}}, \bibinfo{author}{\bibfnamefont{B.}~\bibnamefont{D’Anjou}}, \bibnamefont{and} \bibinfo{author}{\bibfnamefont{A.}~\bibnamefont{Blais}}, \bibinfo{journal}{Physical Review X} \textbf{\bibinfo{volume}{14}}, \bibinfo{pages}{041023} (\bibinfo{year}{2024}), ISSN \bibinfo{issn}{2160-3308}, \urlprefix\url{https://link.aps.org/doi/10.1103/PhysRevX.14.041023}.

\bibitem[{\citenamefont{Sank et~al.}(2016)\citenamefont{Sank, Chen, Khezri, Kelly, Barends, Campbell, Chen, Chiaro, Dunsworth, Fowler et~al.}}]{sank_measurement-induced_2016}
\bibinfo{author}{\bibfnamefont{D.}~\bibnamefont{Sank}}, \bibinfo{author}{\bibfnamefont{Z.}~\bibnamefont{Chen}}, \bibinfo{author}{\bibfnamefont{M.}~\bibnamefont{Khezri}}, \bibinfo{author}{\bibfnamefont{J.}~\bibnamefont{Kelly}}, \bibinfo{author}{\bibfnamefont{R.}~\bibnamefont{Barends}}, \bibinfo{author}{\bibfnamefont{B.}~\bibnamefont{Campbell}}, \bibinfo{author}{\bibfnamefont{Y.}~\bibnamefont{Chen}}, \bibinfo{author}{\bibfnamefont{B.}~\bibnamefont{Chiaro}}, \bibinfo{author}{\bibfnamefont{A.}~\bibnamefont{Dunsworth}}, \bibinfo{author}{\bibfnamefont{A.}~\bibnamefont{Fowler}}, \bibnamefont{et~al.}, \bibinfo{journal}{Physical Review Letters} \textbf{\bibinfo{volume}{117}}, \bibinfo{pages}{190503} (\bibinfo{year}{2016}), ISSN \bibinfo{issn}{0031-9007, 1079-7114}, \urlprefix\url{https://link.aps.org/doi/10.1103/PhysRevLett.117.190503}.

\bibitem[{\citenamefont{Ferrari et~al.}(2025)\citenamefont{Ferrari, Gravina, Eeltink, Scarlino, Savona, and Minganti}}]{ferrari_dissipative_2025}
\bibinfo{author}{\bibfnamefont{F.}~\bibnamefont{Ferrari}}, \bibinfo{author}{\bibfnamefont{L.}~\bibnamefont{Gravina}}, \bibinfo{author}{\bibfnamefont{D.}~\bibnamefont{Eeltink}}, \bibinfo{author}{\bibfnamefont{P.}~\bibnamefont{Scarlino}}, \bibinfo{author}{\bibfnamefont{V.}~\bibnamefont{Savona}}, \bibnamefont{and} \bibinfo{author}{\bibfnamefont{F.}~\bibnamefont{Minganti}}, \bibinfo{journal}{Phys. Rev. Res.} \textbf{\bibinfo{volume}{7}}, \bibinfo{pages}{013276} (\bibinfo{year}{2025}), \urlprefix\url{https://link.aps.org/doi/10.1103/PhysRevResearch.7.013276}.

\bibitem[{\citenamefont{Cohen et~al.}(2023)\citenamefont{Cohen, Petrescu, Shillito, and Blais}}]{cohen_reminiscence_2022}
\bibinfo{author}{\bibfnamefont{J.}~\bibnamefont{Cohen}}, \bibinfo{author}{\bibfnamefont{A.}~\bibnamefont{Petrescu}}, \bibinfo{author}{\bibfnamefont{R.}~\bibnamefont{Shillito}}, \bibnamefont{and} \bibinfo{author}{\bibfnamefont{A.}~\bibnamefont{Blais}}, \bibinfo{journal}{PRX Quantum} \textbf{\bibinfo{volume}{4}}, \bibinfo{pages}{020312} (\bibinfo{year}{2023}), \urlprefix\url{https://link.aps.org/doi/10.1103/PRXQuantum.4.020312}.

\bibitem[{\citenamefont{Arfken et~al.}(2013)\citenamefont{Arfken, Weber, and Harris}}]{arfken_mathematical_2013}
\bibinfo{author}{\bibfnamefont{G.~B.} \bibnamefont{Arfken}}, \bibinfo{author}{\bibfnamefont{H.~J.} \bibnamefont{Weber}}, \bibnamefont{and} \bibinfo{author}{\bibfnamefont{F.~E.} \bibnamefont{Harris}}, \emph{\bibinfo{title}{Mathematical Methods for Physicists: A Comprehensive Guide}} (\bibinfo{publisher}{Academic Press}, \bibinfo{address}{Waltham, MA}, \bibinfo{year}{2013}), \bibinfo{edition}{7th} ed., ISBN \bibinfo{isbn}{9780123846549}.

\bibitem[{\citenamefont{Bultink et~al.}(2018)\citenamefont{Bultink, Tarasinski, Haandbæk, Poletto, Haider, Michalak, Bruno, and DiCarlo}}]{bultink_general_2018}
\bibinfo{author}{\bibfnamefont{C.~C.} \bibnamefont{Bultink}}, \bibinfo{author}{\bibfnamefont{B.}~\bibnamefont{Tarasinski}}, \bibinfo{author}{\bibfnamefont{N.}~\bibnamefont{Haandbæk}}, \bibinfo{author}{\bibfnamefont{S.}~\bibnamefont{Poletto}}, \bibinfo{author}{\bibfnamefont{N.}~\bibnamefont{Haider}}, \bibinfo{author}{\bibfnamefont{D.~J.} \bibnamefont{Michalak}}, \bibinfo{author}{\bibfnamefont{A.}~\bibnamefont{Bruno}}, \bibnamefont{and} \bibinfo{author}{\bibfnamefont{L.}~\bibnamefont{DiCarlo}}, \bibinfo{journal}{Applied Physics Letters} \textbf{\bibinfo{volume}{112}}, \bibinfo{pages}{092601} (\bibinfo{year}{2018}), ISSN \bibinfo{issn}{0003-6951, 1077-3118}, \urlprefix\url{https://pubs.aip.org/apl/article/112/9/092601/36013/General-method-for-extracting-the-quantum}.

\end{thebibliography}
